\documentclass[a4paper,10pt]{article}
\usepackage[utf8x]{inputenc}

\usepackage{graphicx}  
\usepackage{amsmath}   
\usepackage[compress]{cite}
\usepackage{amssymb}   
\usepackage[active]{srcltx}

\def\eq#1{{Eq.~(\ref{#1})}}

\newcommand{\LL}{Lanczos-Lovelock}

\newcommand{\df}{\delta}

\def\cc{{cosmological\ constant}}

\title{Distribution function  of the Atoms of Spacetime and the Nature of Gravity}
\author{T. Padmanabhan\\
IUCAA, Pune University Campus,\\
  Ganeshkhind, Pune- 411 007.\\
{\small{email: paddy@iucaa.in}}
}
\date{ }

\begin{document}

\maketitle

\begin{abstract}
The fact that the equations of motion for matter remain invariant when a constant is added to the Lagrangian suggests postulating that the field equations of gravity should also respect this symmetry. This principle implies that: (1) the metric cannot be varied in any extremum principle to obtain the field equations; and (2) the stress-tensor of matter should appear in the variational principle through the combination $T_{ab}n^an^b$ where $n_a$ is an auxiliary null vector field, which could be varied to get the field equations. This procedure  selects naturally the \LL\ models of gravity in $D$-dimensions and Einstein's theory in $D=4$. Identifying $n_a$ with the normals to the null surfaces in the spacetime \textit{leads to} a thermodynamic interpretation for gravity, in the macroscopic limit.
Several
geometrical variables and the equation describing the spacetime evolution acquire a thermodynamic interpretation. 

Extending these ideas one level deeper, we can obtain this variational principle from a distribution function for the ``atoms of spacetime'', which counts the number of microscopic degrees of freedom of the geometry. This is based on the curious fact that the renormalized spacetime endows each event with zero volume, but finite area!
 \end{abstract}
\tableofcontents

\section{Gravity: An Emergent Phenomenon}\label{sec:gravemerge}

While the difference between a hot body and a cold one was known even to the cavemen, physicists struggled for centuries to understand the nature of heat \cite{tpr:heat}. It was known to them that a macroscopic system like, for example, a gas can be studied by introducing several thermodynamic variables (like temperature, entropy, \textit{etc.}), but for a very long time, they did not know what these variables really meant. The breakthrough came with the work of Boltzmann, who essentially said: ``If you can heat it, it has microscopic degrees of freedom''. Before this idea was accepted, a gas or a fluid was thought of as a continuum all the way down to the smallest scales, and the notion of heat and temperature were superimposed on it, in a rather \textit{ad hoc} manner. Boltzmann introduced a paradigm shift in which matter was treated as discrete at small scales and the thermal phenomena were related to the (suitably averaged) mechanical attributes of these discrete degrees of freedom.

This paradigm shift is profound. It stresses that \textit{the existence of microscopic degrees of freedom leaves a tell-tale signature even at the largest macroscopic scales, in the form of temperature and heat}. One could have guessed that a glass of water {must be} made of discrete microscopic degrees of freedom just from the fact that it can be heated, without probing it at Angstrom scales, even though it actually took centuries for physicists to recognize that temperature and heat provide a direct link between microscopic and macroscopic phenomena. In fact, a relation like $Nk_B=E/[(1/2)T]$ directly counts the microscopic degrees of freedom, $N$, in terms of macroscopic variables $E$ and $T$!

Mathematically, one key variable in thermodynamics, which was absent in the Newtonian mechanics of point particles, is the heat content $TS$ of the matter, which is the difference $(F-E)$ between the free energy and the internal energy of the system. In terms of densities, the heat density is $Ts = P+\rho$, where $s$ is the entropy density, $\rho$ is the energy density and $P$ is the pressure. (This is the Gibbs--Duhem relation for systems with zero chemical potential in which we will be interested).

Proceed now from normal matter to spacetime. Work done in the last several decades \cite{A1,A2,A3,A4,A5,A6,A7,A8} shows that spacetimes, due to the existence of null surfaces, which block information from a certain class of observers, also possess a heat density $Ts$. The emergent gravity paradigm \cite{A9,A11} builds upon this fact and treats the gravitational field equations as analogous to the equations of fluid dynamics or elasticity. There is a considerable amount of {internal} evidence in the structure of gravitational theories, much more general \cite{A13,A34} than Einstein's theory, to indicate that this is a correct and useful approach to pursue. This review explores several aspects of this approach and extends the ideas to a deeper level.

\section{Scope, Structure and Features of this Review}

As will become clear soon, it is possible to associate a temperature and entropy density with every event in the spacetime just as one could have done so for a glass of water. 
On the other hand, one traditionally described the dynamics of spacetime through some field equation for gravity, because Einstein told us that gravity is nothing but the curvature of spacetime. If we take both of these results seriously, we are led to the following conclusions and results described in this review:

\begin{enumerate}

 \item The Boltzmann principle suggests that if spacetime can be hot, it must have a microstructure. What is more, we should be able to count the atoms of spacetime without having the technology to do Planck-scale experiments, just as Boltzmann could guess the existence of atoms of matter without doing Angstrom-scale experiments. We would then expect a relation like $Nk_B=E/[(1/2)T]$ to exist for the spacetime. We will see in Section~\ref{sec:avogadro} that this is indeed the case.

\item 

If the spacetime is analogous to a fluid made of atoms, the gravitational field equations must have the same conceptual status as the equations describing fluid mechanics. Hence, we should be able to derive them from a purely-thermodynamic variational principle. Just as in the case of matter, such a variational principle \cite{A14,A16} will be a phenomenological input when we approach it from the macroscopic side.

Further, we should be able to write the field equation in a purely \textit{thermodynamic} language rather than in the (conventional) \textit{geometrical} language \cite{A19,A20,A21,A22}. Consequently, we would expect several variables, which are usually considered geometrical, to have an underlying thermodynamic interpretation. We will describe these features in Sections~\ref{sec:elegantgravdyn} and~\ref{sec:geotherm}.

\item The discreteness of normal matter is usually taken into account in the kinetic theory by introducing a distribution function $f(x^i, p_i)$, such that $dN = f(x^i, p_i) d^3x d^3p$ counts the number of atoms in a phase volume. Such a description recognizes the discreteness, but works at scales such that the volume $d^3x$ is large enough to, say, contain a sufficient number of atoms. We can develop (see Sections~\ref{sec:gravheatden} and~\ref{sec:eventarea}) a similar concept for the spacetime that recognizes the discreteness at the Planck scale and yet allows the use of continuum mathematics to describe the phenomena. This provides a deeper level of description of spacetime, such that the thermodynamic variational principle, mentioned in Item (2) above, can be obtained from it.

\item Such a reformulation of spacetime dynamics as thermodynamics should provide us with insights into some of the problems of the standard formulation, like for example, the cosmological constant, spacetime singularities, \textit{etc}. This goes beyond describing what is known in a \textit{new language} and should lead to \textit{new results} \cite{C8,C9}. I will describe in Sections~\ref{sec:eventarea} and~\ref{sec:summary} how this approach leads to a new perspective on cosmology and allows us to {predict the numerical value} of the \cc!

\end{enumerate}

There exists a fair amount of previous work (cited above) that shows that the emergent gravity paradigm does achieve 1, 2 and 4 above. In Sections~\ref{sec:gravbricks} and \ref{sec:geotherm}, we will review these developments, highlighting some recent results. The main thrust of this article, however, will be to describe (Sections~\ref{sec:gravheatden}--\ref{sec:summary}) the first glimpses of a viable microscopic model, related to Item (4) above, and to explain how one could possibly recover spacetime thermodynamics as a limit of the statistical mechanics of the atoms of spacetime.\footnote{\textit{Notation:} The signature is $(-,+,+,+, ...)$. The Latin letters run over all of the spacetime indices $(0,1,2,....d-1)$; the Greek letters over the spatial indices $(1,2,....d-1)$; and the uppercase Latin letters, $A,B,C,\ldots$, run over a co-dimension two surface when appropriate. We set $\hbar=1,c=1$ and $16\pi G=1$ for the most part of our discussion (occasionally, when we use the $G=1$ units, it will be mentioned specifically). Einstein's field equations will then take the form $2G_{ab}=T_{ab}$.}

\section{Building Gravity: Brick by Brick}\label{sec:gravbricks}

I will begin by describing the logical structure behind a first-principle approach, which obtains the spacetime dynamics as an emergent phenomenon, working from the macroscopic side. 

To do this, it is convenient to separate the kinematic (``how gravity makes the matter move'') and dynamic (``how matter makes the spacetime curve'') aspects of the gravitational theories. This is important, because there is some amount of emotional resistance in the community to tinkering with general relativity, given its elegance and beauty. However, what is not often recognized (or stressed in the text books) is that \textit{all} of the elegance of general relativity is confined to its {kinematic part}, which describes gravity as being due to the curvature of spacetime. The dynamics, encoded in the gravitational field equations, has no real elegance and, in fact, does not follow from any beautiful principle analogous to, for example, the principle of equivalence. 
The emergent gravity paradigm retains {all} of the elegance of general relativity by keeping its kinematic structure intact; further, it provides a nice thermodynamic underpinning to describe the dynamics. 
In Sections~\ref{sec:elegantgravkin} and \ref{sec:elegantgravdyn}, I will describe how this comes about.

\subsection{The Elegance of Gravitational Kinematics}\label{sec:elegantgravkin}

Judicious use of the principle of equivalence tells us that gravity \textit{is} geometry and can be described by a metric $g_{ab}$ of the curved spacetime. Further, the principle of general covariance insists on the democratic treatment of all observers in the spacetime. By abandoning any special form of the pre-geometric metric (like the $\eta_{ab}$ of special relativity), we accept the fact that one can no longer think of a part of $g_{ab}$ as arising due to acceleration (\textit{i.e.}, coordinate choice) and a part as arising due to genuine curvature. These principles also provide us with a procedure to describe the influence of spacetime geometry on matter fields: we invoke the standard laws of special relativity (SR) in a freely-falling frame (FFF), rewrite them in a generally covariant language valid in arbitrary curvilinear coordinates and postulate that the same form should hold, even in a curved spacetime. As a consequence, the energy momentum tensor $T^a_b$ for the matter (known from SR) will satisfy the equation: 
\begin{equation}
\nabla_a T^a_b = 0 
\label{divT}
\end{equation} 
in curvilinear coordinates in SR and, hence, should also hold in arbitrary curved geometry. Generically, this equation will give the equations of motion for matter in the presence of gravity. In our approach, the matter sector will be described by a $T^a_b$, which satisfies \eq{divT}, rather than by an action, \textit{etc}.

It is also straightforward to conclude from  \eq{divT}, applied  to the  light rays, that they will bend in the presence of gravity;   hence the causal structure of the spacetime will now be determined by the gravitational field. In particular, it is easy to construct observers (\textit{i.e.}, timelike congruences) in any spacetime such that part of the spacetime will be inaccessible to them.\footnote{I stress that (a) this is a purely kinematic feature and (b) it is \textit{always} observer dependent. For example, (i) such observers exist even in flat spacetime and (ii) in the case of, say, a black hole spacetime, an observer freely falling into the black hole and the one who is stationary outside, will access different regions of spacetime.} A generic example of such  observers is provided by the local Rindler observers \cite{A35} constructed as follows: 
In a region around any event $\mathcal{P}$, introduce the FFF with coordinates $(T, \mathbf{X})$. Boost from the FFF to a local Rindler frame (LRF) with coordinates $(t,\mathbf{x})$ constructed using some acceleration $a$, through the transformations: $X=x\cosh (at), T=x\sinh (at)$. There will be a null surface passing though $\mathcal{P}$, which gets mapped to the $X=T$ surface in the FFF; this null surface will now act as a patch of horizon to the $x=$ constant Rindler observers.

This construction leads to the most beautiful result \cite{A5,A6} we have obtained so far by combining the principles of general relativity and quantum field theory: the local vacuum state, defined by the freely-falling observers around an event, will appear as a thermal state to the local Rindler observer with the temperature: 
\begin{equation}
 k_BT = \left(\frac{\hbar}{c}\right) \left(\frac{a}{2\pi}\right)
\end{equation} 
where $a$ is the acceleration of the local Rindler observer, which can be related to other geometrical variables of the spacetime in different contexts. 
This Davies--Unruh temperature tells us that around \textit{any} event, in \textit{any} spacetime, there exists a class of observers who will perceive the spacetime as hot. This fact will play a crucial role in our discussion.

There are a couple of related results that we will use later on, which are worth recalling at this stage. The~first is the relation between Euclidean spacetime and the temperature introduced above. The~mapping, from the FFF to the LRF, $X=x\cosh at,\ T=x\sinh at$, has the Euclidean continuation (under \mbox{$iT = T_E, \ it = t_E$}) given by 
$X=x\cos at_E,\ T_E=x\sin at_E$. This, in turn, maps a pair of null surfaces $X^2 - T^2 =0$ to the single point in the Euclidean origin given by $X^2 + T_E^2 =0$. Approaching the origin of the Euclidean sector, therefore, corresponds to approaching the null surface in the original spacetime as a limit. We will make use of this fact later on.

The second result \cite{A35} is related to the energy flow associated with the matter that crosses the null surface, as viewed from the FFF. A local Rindler observer will see that the matter takes a very long time to cross the local Rindler horizon, thereby allowing for thermalization to take place. (This is similar to the fact that, as seen by the outside observer, matter takes infinite time to cross the black hole horizon). Since the local Rindler observer attributes a temperature $T$ to the horizon, she will interpret the energy associated with the matter that crosses the null surface (asymptotically) as some amount of energy $\Delta E$ being dumped on a \textit{hot} surface, thereby contributing a \textit{heat} content $\Delta Q=\Delta E$. This quantity can be computed as follows:

We choose an FFF around any given spacetime event $\mathcal{P}$ and construct an LRF. The LRF provides us with an approximate Killing vector field $\xi ^{a}$, generating boosts, which coincides with the null normal $\ell ^{a}$ at the null surface. The heat current arises from the boost energy current $T_{ab}\xi ^{b}$ of matter. Therefore, the total heat energy dumped on 
the null surface will be:
\begin{align}\label{Paper06_New_11}
 Q_{m}=\int \left(T_{ab}\xi ^{b}\right)d\Sigma ^{a}=\int T_{ab}\xi ^{b}\ell ^{a}\sqrt{\gamma}d^{2}x d\lambda
=\int T_{ab}\ell ^{b}\ell ^{a}\sqrt{\gamma}d^{2}x d\lambda
\end{align}
where we have used the fact that $\xi ^{a} \to \ell ^{a}$ on the null surface. 
Since the parameter $\lambda $ (defined through $\ell^a = dx^a/d\lambda$) is similar to a time coordinate,
we can also define a heating rate: 
\begin{equation}
 \frac{dQ_{m}}{d\lambda}=\int T_{ab}\ell ^{b}\ell ^{a}\sqrt{\gamma}d^{2}x
 \end{equation} 
and a heating rate density per unit proper area of the surface:
\begin{equation}
 \mathcal{H}_m[\ell_a]\equiv \frac{dQ_{m}}{\sqrt{\gamma}d^{2}xd\lambda}=T_{ab} \ell^a\ell^b
\label{hmatter} 
\end{equation}
so that the heat transferred by matter is obtained by integrating $\mathcal{H}_m$ with the integration measure $\sqrt{\gamma}d^{2}xd\lambda$ over the null surface generated by the null congruence $\ell ^{a}$, parametrized by $\lambda$.
We will simply call $\mathcal{H}_m$ the heat density (energy per unit area per unit time) of the null surface, contributed by matter crossing a local Rindler horizon, as interpreted by the local Rindler observer.
There are two features that are noteworthy regarding this heat density.

\begin{itemize}

 \item If we add a constant to the matter Lagrangian (\textit{i.e.}, $L_{m} \to L_{m} + $ constant), the $T^a_{b}$ changes by $T^a_b \to T^a_b + $ (constant) $\delta^a_b$. The heat density, defined by Equation~(\ref{hmatter}) remains invariant under this transformation. 

\item The heat density vanishes if $T^a_b \propto \delta^a_b$. Therefore, \textit{the cosmological constant has zero heat density}, though it has non-zero energy density. 
(In fact, for an ideal, comoving fluid, $T_{ab} \ell^a\ell^b = (\rho + P)$, and hence, the heat density vanishes only for the \cc\ with equation of state $\rho=-P$.)

\end{itemize}
We will have occasion to use these facts later on.

\subsection{Restoring Elegance to Gravitational Dynamics}\label{sec:elegantgravdyn}

The next task is to obtain the field equations describing the evolution of spacetime geometry. In the conventional approach, there is no simple guiding principle that will allows us to do this, and it ultimately reduces to certain assumptions of simplicity. I will now show how it is possible to approach gravitational dynamics using a guiding principle, which turns out to be as powerful as the principle of equivalence~\cite{A19,A11}.

Recall that the equations of motion for matter, obtained from an action principle, remain invariant if we add a constant to the matter Lagrangian, \textit{i.e.}, under $L_{m} \to L_{m} + $ constant.\footnote{To be precise, there is some subtlety if supersymmetry is an unbroken symmetry; since we have no evidence for supersymmetry anyway, I will not discuss this issue.} Mathematically, this is a trivial consequence of the fact that the Euler equations only care about the \textit{derivatives} of the Lagrangian. Physically, this encodes the principle that the zero level of energy density does not affect dynamics. It~seems reasonable to postulate that the gravitational field equations should not break this symmetry, which is already present in the matter sector. Since $T_{ab}$ is the most natural source for gravity (as can be argued from the principle of equivalence and considerations of the Newtonian limit), we demand that:

\begin{itemize}

 \item[$\blacktriangleright$] The variational principle that determines the dynamics of spacetime must be invariant under the change $T^a_b \to T^a_b + $ (constant) $\delta^a_b$.

\end{itemize}

This principle immediately rules out the possibility of varying the metric tensor $g_{ab}$ in a covariant, local, action principle to obtain the field equations! It can be easily proven \cite{C7} that if (i) the action is obtained from a local, covariant Lagrangian integrated over a region of spacetime with the standard measure $\sqrt{-g}\, d^4x$ and (ii) the dynamical equations are obtained by the unrestricted variation\footnote{The second condition rules out unimodular \cite{C18,C19,C20} theories and their cousins, in which one varies the metric keeping $\sqrt{-g}$ fixed; I do not think we have good physical motivation for this approach.} of the metric in the action, then the field equations \textit{cannot} remain invariant under $T^a_b \to T^a_b + $ (constant) $\delta^a_b$.  In fact, $L_{m} \to L_{m} + $ constant is no longer a symmetry of the action if the metric is treated as the dynamical variable. Therefore, any variational principle we want to work with cannot use $g_{ab}$ as a dynamical variable. 
This fact, in turn, raises two issues:

(1) Normally, you will vary some variables $q_A$ in an action to obtain equations of motion for the \textit{same} variables $q_A$. We, of course, want the dynamical equation to still describe the evolution of $g_{ab}$, but we have just concluded that we cannot vary $g_{ab}$ in any variational principle! How is this possible?

(2) In the conventional approach, we vary the metric in the \textit{matter} Lagrangian to obtain $T_{ab}$ as the source. Since we are not varying $g_{ab}$, but still want $T_{ab}$ to be the source, it is necessary to have $T_{ab}$ explicitly included in the variational principle. Therefore, we want the variational principle to \textit{depend} on $T_{ab}$ and, yet, \textit{be invariant} under $T^a_b \to T^a_b + $ (constant) $\delta^a_b$! How can this be done?

The answers to these two questions are closely related. The combination $T_{ab} n^a n^b$, where $n_a$ is any null vector, is obviously invariant under the shift $T^a_b \to T^a_b + $ (constant) $\delta^a_b$. Therefore, if the variational principle depends on $T_{ab}$ only through the combination $T_{ab} n^an^b$, the requirement in (2) above is automatically satisfied.\footnote{We want to introduce a minimum number of auxiliary variables. In a $d$-dimensional spacetime, the null vector with $(d-1)$ degrees of freedom is the minimum. In contrast, if we use, say, a combination $T^{ab}V_{ab}$ with a symmetric traceless tensor $V_{ab}$, to maintain the invariance under $T^a_b \to T^a_b + $ (constant) $\delta^a_b$, then we would have introduced $(1/2)d(d+1)-1$ degrees of freedom; in $d=4$, this introduces nine degrees of freedom, equivalent to introducing three null vectors rather than one.}
This suggests using a variational principle that extremizes a functional defined by:
\begin{equation}
Q_{\rm tot}\equiv \int dV (\mathcal{H}_m+\mathcal{H}_g); \qquad \mathcal{H}_m[n_a] \equiv T_{ab} n^a n^b
\label{Qtot}
\end{equation} 
where $\mathcal{H}_g$ is the corresponding contribution from gravity, which
is yet to be determined, and $dV$ is the proper measure for integration over a suitable region of spacetime, which is also currently left unspecified. This approach introduces an arbitrary null vector $n_a$ into the variational principle, which at this stage, is just an auxiliary field. However, since no null vector is special, the extremum should hold for all null vectors, which requires us to vary $n_a$ in \eq{Qtot} and demand that the resulting equations hold for \textit{all} $n_a$ at a given event. This should lead to a constraint on the background metric $g_{ab}$, which will determine the dynamics of spacetime. 
If we can find such a $\mathcal{H}_g$, we would have taken care of the issue raised in (1) above, as well.

Therefore, we need to find a suitable functional $\mathcal{H}_g[n_a, g_{ab}]$ of $n_a(x), g_{ab}(x)$, such that the extremum condition $\delta Q_{tot}/\delta n_a=0$, for all null vectors $n_a$ at a given event $\mathcal{P}$, leads to sensible equations for the evolution of $g_{ab}$. Since $Q_{tot}$ is invariant under $T^a_b \to T^a_b + $ (constant) $\delta^a_b$,
the source that appears in the field equation must respect this symmetry. Therefore,
we would expect the equations of motion to be algebraically equivalent~to: 
\begin{equation}
2E^a_b=T^a_b+\Lambda \delta^a_b 
\label{eab}              
\end{equation} 
Here, $\Lambda$ is an \textit{undetermined integration constant}, which will allow us to absorb the constant in the shift $T^a_b \to T^a_b + $ (constant) $\delta^a_b$, while $E^a_b$ is constructed from $g_{ab}$ and its derivatives and must satisfy $\nabla_a E^a_b=0$ identically for consistency.

By very construction, the cosmological constant (for which $T_{ab}^{(\Lambda)}n^an^b=0$) cannot appear in the variational principle. At the same time, it arises as an integration constant in \eq{eab}, and we need a \textit{further} principle to fix its value once and for all. 
Therefore, the microscopic theory, \textit{viz}. the statistical mechanics of atoms of spacetime, should lead to: 
\begin{itemize}

 \item The explicit form of $\mathcal{H}_g[n_a, g_{ab}]$ in the thermodynamic limit. 

 \item A procedure to determine the value of the cosmological constant in our universe. 

\end{itemize}
I will describe later on (see Section~\ref{sec:kineticsast}) how one could attempt to model such a microscopic theory that will satisfy both of these criteria, but first, I will show how one can obtain the form of $\mathcal{H}_g[n_a, g_{ab}]$ working downward from the macroscopic description. 

Everything works out fine \cite{A14,A16} if we take $\mathcal{H}_g$ to be a quadratic in $\nabla_an_b$ of the form:
\begin{equation}
\mathcal{H}_g= -\left(\frac{1}{16\pi L_P^2}\right) (4P^{ab}_{cd}\nabla_an^c\nabla_bn^d)
\label{hgrav}
\end{equation}
where $L_P^2$ is an arbitrary constant at this stage, with the dimensions of area (this gives $\mathcal{H}_g$ the dimension $L^{-4}$ as required).
Demanding that $\delta Q_{tot}/\delta n_a=0$ for all null vectors $n_a$ at a given event should lead to an equation for background geometry allows us to fix the form of $P^{ab}_{cd}$.
We find that:
\begin{equation}\label{Paper06_SecLL_07}
P^{ab}_{cd} \propto \delta ^{aba_{2}b_{2}\ldots a_{m}b_{m}}_{cdc_{2}d_{2}\ldots c_{m}d_{m}}
R^{c_{2}d_{2}}_{a_{2}b_{2}}\ldots R^{c_{m}d_{m}}_{a_{m}b_{m}}
\end{equation}
where $\delta ^{aba_{2}b_{2}\ldots a_{m}b_{m}}_{cdc_{2}d_{2}\ldots c_{m}d_{m}}$ is the totally-antisymmetric $m$-dimensional determinant tensor. 
If we now extremize $Q_{tot}$ in \eq{Qtot}, using this $P^{ab}_{cd}$ in the expression for $\mathcal{H}_g$ in \eq{hgrav}, 
we get the field equations of (what is known as) the \LL\ model \cite{A13,A14,A16}, given by:
\begin{equation}
E^{a}_{b}\equiv P^{ai}_{jk}R_{bi}^{jk}-\frac{1}{2}\delta^{a}_{b}\mathcal{R}
=(8\pi L_P^2)T^{a}_{b}+\Lambda\delta^a_b       
\end{equation} 
where $E^{a}_{b}$ and
$
m\mathcal{R}\equiv P^{ab}_{cd}R_{ab}^{cd}
$ 
are the generalizations of the Einstein tensor and the Ricci scalar.\footnote{It is possible to prove that $E_{ab}$ is \textit{symmetric} \cite{A40} and $\nabla_aE^a_b =0$, so that everything is consistent. Further, the variational principle works when $dV$ in \eq{Qtot} is the integration measure on the spacetime or on a suitable null surface with $n_a$ as the normal.}
These models \cite{A37,A38,A39} have the curious, and unique, feature that, even though the Lagrangians describing them, in the conventional approach, are $m$-th degree polynomials in the curvature tensor, the resulting field equations are still second order in $g_{ab}$!

In $d=4$ dimensions, $P^{ab}_{cd}$ reduces to the determinant tensor given by 
$
P^{ab}_{cd}=(1/2)(\delta^a_c\delta^b_d-\delta^b_c\delta^a_d)
$.
The resulting equation for $g_{ab}$ is identical to Einstein's equations with an undetermined cosmological~constant: 
\begin{equation}
G^a_b=(8\pi L_P^2)T^a_b+\Lambda \delta^a_b                   
\end{equation} 
which has the structure in \eq{eab}, as expected. 

The expression for $P^{ab}_{cd}$ determines the entropy density of horizons (corresponding to the Wald entropy) in the resulting theory through the expression \cite{A8,A13}:
\begin{equation}
s=-\frac{1}{8} \sqrt{\gamma} P^{abcd}\epsilon _{ab}\epsilon _{cd}
\label{entrophys}
\end{equation}
(where $\epsilon_{ab}$ is the binormal to the horizon surface)
which, of course, reduces to $\sqrt{\gamma}/4$ if we choose $P^{ab}_{cd}=(1/2)(\delta^a_c\delta^b_d-\delta^b_c\delta^a_d)$, appropriate for the Einstein gravity. Thus, the specification of horizon entropy specifies the $P^{ab}_{cd}$ and selects the corresponding \LL\ model. In the case of normal matter, we know that two different bodies, say, a glass of water and a metal rod, can be kept at the same temperature; so, the temperature of a material is purely kinematic and contains no structural information. On the other hand, the entropy function $S(E,V)$ will be completely different for water and the metal rod at the same temperature; specifying it will allow us to describe the structure of the material. Similarly, the temperature of the spacetime, as we saw before, is purely kinematic, but specifying the form of horizon entropy in \eq{entrophys}, specifies the dynamics of the theory.

So far, we have not specified the physical nature of null vector field $n^a$ nor the physical interpretation of $\mathcal{H}_g$ or $\mathcal{H}_m$. We, however, know from Equations~(\ref{Paper06_New_11}) and (\ref{hmatter}) that the combination $T_{ab}n^an^b$ has a physical interpretation (of the heat density contributed by matter to a null surface), if we identify $n^a=\ell^a$, the tangent vector to a null congruence defining a null surface, and 
choose $dV=\sqrt{\gamma}\, d^2xd\lambda$, which is the natural integration measure on the null surface.
The identifications, $n_a\to \ell_a$ with $\mathcal{H}_m[n]\to \mathcal{H}_m[l]$, in turn, imply that $\mathcal{H}_g[\ell_a]$ should be interpreted as the corresponding quantity, \textit{viz}. the heat density contributed by gravity to the null surface. Thus, our guiding principle, that the field equations should be invariant under $T^a_b \to T^a_b + $ constant $\delta^a_b$, tells us that the variational principle extremizes the total heat density (since we know the interpretation of $\mathcal{H}_m$ for matter), thereby leading to a direct thermodynamic interpretation to the variational principle based on: 
\begin{equation}
 Q_{\rm tot} \equiv \int \sqrt{\gamma}\, d^2xd\lambda\, (\mathcal{H}_g[\ell] +\mathcal{H}_m[\ell]) 
\end{equation} 

Since $P^{ab}_{cd}$ is related to the entropy of the horizons in the resulting theory, it is no surprise that the on-shell value of $Q_{\rm tot}$ is closely related to the entropy of null surfaces. We can show \cite{SCTPnull} in general relativity, for example, that the on-shell value is:
\begin{equation}
 Q_{\rm tot}^{({\rm on-shell})} = Q(\lambda_2)-Q(\lambda_1)
 \label{eqnx2}
\end{equation} 
with: 
\begin{equation}
 Q(\lambda)= \int \frac{\sqrt{\gamma} \, d^2x}{4L_P^2}\, k_B T_{\rm loc} = \int d^2x (T_{\rm loc}\, s)
\end{equation} 
where $T_{\rm loc}$ is the Davies--Unruh temperature attributed to the null surface by appropriate local Rindler observers and $s= (\sqrt{\gamma}/4L_P^2)$ is the entropy density in \eq{entrophys} for general relativity (the interpretation in \eq{eqnx2} works for all \LL\ models if we use the $s$ in \eq{entrophys}).

It is also possible to provide a direct physical meaning to $L_P^2$. This is most easily found from rewriting \eq{eqnx2} as:
\begin{equation}
 2Q_{\rm tot}^{({\rm on-shell})} = E_{\rm sur}(\lambda_2) - E_{\rm sur}(\lambda_1)
 \label{eqnx1}
\end{equation} 
with:
\begin{equation}
E_{\rm sur}(\lambda) = \int \frac{\sqrt{\gamma} \, d^2x}{L_P^2} \left( \frac{1}{2}k_B T_{\rm loc}\right) 
=\frac{1}{2} N_{\rm sur} (k_BT_{\rm avg})
\label{equiE}
\end{equation}
where $T_{\rm avg}$ is the average of $T_{\rm loc}$ over the surface and $N_{\rm sur} =(A_{\rm sur}/L_P^2)$ is the number of surface degrees of freedom \cite{A17,A18} if we attribute one degree of freedom to each cell of area $L_P^2$.
\textit
{This provides the physical meaning of the fundamental constant $L_P^2$ we have introduced as a quantum of area}; viz., the number of microscopic degrees of freedom associated\footnote{One can, of course, rescale $(1/2)k_BT\to(\nu/2)k_BT,N_{sur}\to A_{sur}/\nu L_P^2$ without changing the result; we have chosen $\nu=1$.} with an area $A$ is $A/L_P^2$.
Therefore, the physical meaning of $Q_{\rm tot}$, used in our variational principle, is reiterated by its on-shell value.\footnote{The relative factor two in the left-hand sides of Equations~(\ref{eqnx1}) and (\ref{eqnx2}) is not \textit{ad hoc} and, in fact, helps to solve a long-standing problem in general relativity related to a factor two in the definition of Komar mass; see, e.g., \cite{C4}; I will not discuss it here.}

The following point, however, needs to be stressed. Eventually, one would like to obtain such a thermodynamic variational principle from a deeper, microscopic consideration. All that we require in such a derivation is that (i) \textit{some} auxiliary null vector field $n_a$ should arise in the microscopic theory and (ii) should lead to $\mathcal{H}_g [n_a]$ with the correct structure. If we identify this $n_a$ with the normals to the null surfaces, we get the correct field equations in the macroscopic limit. However, at a fundamental level, the auxiliary vector field $n^a$ (which could arise in the microscopic physics) and the $\ell^a$ (associated with the null surfaces in the macroscopic limit) are conceptually distinct. I will discuss this in greater detail in Sections~\ref{sec:gravheatden} and \ref{sec:kineticsast}.

The fact that the thermodynamic description transcends general relativity in a unified manner is a feather in the cap for this approach. \textit{In fact, virtually every result in the emergent gravity paradigm obtained for general relativity also holds \cite{A13,A22,A32,A34} for the \LL\ models.} At the same time, the paradigm is quite selective; while it leads to  the \LL\ models with a natural quadratic expression for $Q_{\rm tot}$,  there is no natural generalization to obtain, say, the $f(R)$ models of gravity. The fact that the \LL\ models are the only ones that have 
field equations that are second order in $g_{ab}$ seems to be encoded in this paradigm.
For most of the remaining part of the review, I will work with $d=4$ and general relativity.

The form of $\mathcal{H}_g$ is, of course, not unique, and 
we can add to it any scalar function $f(x)$, possibly built from the metric and other background variables; this will not change the result, because we are varying $\ell_a$ and not $g_{ab}$.
One can also add to it any total derivative of the form $dF/\sqrt{\gamma}d\lambda$, where $F$ can depend on $\ell_a$; such a term will contribute only at the end points $\lambda= \lambda_1,\lambda_2$, where, as usual, we will keep $\ell_a$ fixed. (We can also add a two-divergence $D_Av^A$ in the transverse space, which integrates to zero on $\sqrt{\gamma}d^2x$ integration, and hence is not of much significance). Therefore, a more general form is:
\begin{equation}
 \mathcal{H}_g=f(x)-\left(\frac{1}{16\pi L_P^2}\right) (4P^{ab}_{cd}\nabla_a\ell^c\nabla_b\ell^d)+\frac{1}{\sqrt{\gamma}}\frac{dF}{d\lambda}+D_Av^A
\end{equation} 
This possibility of adding a $(1/\sqrt{\gamma})(dF/d\lambda)$ allows us to rewrite $\mathcal{H}_g$ in a simpler form, which makes the final result obvious. It also helps to separate the contributions that arise even (in, say, a Rindler frame) in flat spacetime from the effects of curvature; in fact, we would expect $\mathcal{H}_g$ to become a total divergence in flat spacetime. I will get back to these aspects
later in Section~\ref{sec:gravheatden}.

There is an important insight we can obtain from this exercise as regards gravity, in spite of the fact that the field equations are the same.
In the Newtonian limit, the gravitational force is now given by: 
\begin{equation}
F=\left(\frac{c^3 L_P^2}{\hbar}\right)\left(\frac{m_1 m_2}{r^2}\right) 
\label{newton}               
\end{equation} 
in terms of the three constants that we have introduced: $c,\hbar,L_P^2$. \textit{You should resist the temptation to write $(c^3L_P^2/\hbar)$ as $G_N$, thereby making $G_N$ independent of $\hbar$!} \eq{newton} tells us that gravity has no classical limit \cite{noclimit}, and the force diverges when $\hbar\to 0$ at finite $L_P^2$, just as all matter collapses when $\hbar\to 0$, because no stable atom can exist. \textit{Gravity is quantum mechanical at all scales.}

To summarize, we have succeeded in obtaining the equations for spacetime evolution, such that: 
(1) The variational principle remains invariant under the shift $T^a_b \to T^a_b + $ (constant) $\delta^a_b$. 
(2) The variational principle is thermodynamic in character and extremizes the heat content of the null surfaces in the spacetime. 
(3) The cosmological constant arises as an integration constant to the equations (and its value needs to be fixed by some further microscopic principle once and for all). The really significant result is:
\begin{itemize}
\item[$\blacktriangleright$] The most natural way of incorporating the fact that gravity is immune to the zero-level of energy \textit{implies}
an emergent, thermodynamic, interpretation for gravity! 
\end{itemize}
This result connects what used to be thought of as two completely separate ideas!

\section{Geometry in the Thermodynamic Language}\label{sec:geotherm}

We have found the dynamical equations for the spacetime, but, as we said earlier (see Item 2 in Section~\ref{sec:gravemerge}), it does not make much sense to use the geometrical language to describe the spacetime evolution if the field equations have the same status as those in other emergent phenomena! 

We saw that the thermodynamic interpretation of geometry relates $L_P^2$ to the degrees of freedom and entropy of null surfaces. This idea will be reinforced when we express the dynamical equations in a thermodynamic language. This has been described in several previous works in this subject cited earlier. For the sake of completeness, I shall review some of the key results, amplifying the conceptual~aspects. 

One way to do this is to introduce \cite{SCTPnull} a conserved vector current $J^a[v]$, which can be defined in terms of an arbitrary vector field $v^a$ in the spacetime. (We will define $J^a[v]$ in the context of 
general relativity, but it can be generalized to all the \LL\ models). This current, when computed for the time evolution vector field $v^a=\xi^a$ in the spacetime, will have a direct thermodynamic significance. 

From any arbitrary vector field $v^{a}$, we can construct\footnote{This happens to be the off-shell version of the standard Noether current; but, the conventional way of deriving it using diffeomorphism invariance of the gravitational action is misleading, because it suggests that $J^a[v]$ has something to do with the action and its symmetries. As we see here, it has nothing to do with either, and its conservation is a rather trivial identity. We will continue to call it the Noether current, but its conservation does not require the nice theorems Emmy Noether proved!} a \textit{conserved} current $J^{a}=\nabla _{b}J^{ab}$
where the antisymmetric tensor $J^{ab}$ is defined as: $ (16\pi L_P^2) J^{ab}=\nabla ^{a}v^{b}-\nabla ^{b}v^{a}$. (The normalization of this current is arbitrary; the introduction of the area $L_P^2$ in the proportionality constant gives it the correct dimension and makes the later results transparent and simple. We will usually work in units with $16\pi L_P^2=1$ and reintroduce it in the final formulas.)  Elementary algebra now leads to the alternative expression:
\begin{equation}
\sqrt{-g}\, J^{a}(v)
=2\sqrt{-g}\, R^{a}_{b}v^{b}+f^{bc}\pounds _{v}N^{a}_{bc}
\label{Paper06_Sec03_Eq02}
\end{equation}
where: 
\begin{align}\label{Paper06_Sec_01_Eq01}
f^{ab}\equiv\sqrt{-g}g^{ab};\qquad N^{c}_{ab}\equiv-\Gamma ^{c}_{ab}+\frac{1}{2}\left(\delta ^{c}_{a}\Gamma ^{d}_{db}+\delta ^{c}_{b}\Gamma ^{d}_{ad}\right)
\end{align}
The individual terms in \eq{Paper06_Sec03_Eq02} are generally covariant, because the Lie derivative of the connection $\pounds_v\Gamma^c_{ab}$, given by
$
\pounds _{v}\Gamma ^{a}_{bc}=\nabla _{b}\nabla _{c}v^{a}+R^{a}_{~cmb}v^{m}
$,
is generally covariant. 

The set $(f^{ab},N^{c}_{ab})$ contains the same amount of information as $(g_{ab},\Gamma^{c}_{ab})$, but has a more direct thermodynamic interpretation \cite{A21}. Let $\mathcal{H}$ be a null surface, which is perceived as a horizon by local Rindler observers who attribute to it a temperature $T$ and entropy density $s=\sqrt{\gamma}/4$.
Then, one can show that:
\begin{itemize}
 \item The combination $ N^{c}_{ab} f^{ab}$, integrated over $\mathcal{H}$ with the usual measure $d^3 \Sigma_c=\ell_c\sqrt{\gamma}d^2xd\lambda$, gives its
heat content; that is:
 \begin{equation}
 \frac{1}{16 \pi L_P^2}\int d^3 \Sigma_c (N^{c}_{ab} f^{ab}) =\int d\lambda\ d^2x\ T s
 \end{equation} 

 \item Consider the metric variations $\delta f$ that preserve the null surface. Remarkably enough, the combinations $f\delta N$ and $N\delta f$ correspond to the variations $s\delta T$ and $T\delta s$, when integrated over the null surface. That is:
 \begin{eqnarray}
\frac{1}{16 \pi L_P^2}\int d^3 \Sigma_c(N^{c}_{ab}\df f^{ab})&=& \int d\lambda\ d^2x\ T \df s; 
\label{stsdt0}\\
\frac{1}{16 \pi L_P^2}\int d^3 \Sigma_c (f^{ab}\df N^{c}_{ab})&=& \int d\lambda\ d^2x\ s \df T
\label{stSdT}
\end{eqnarray}

Therefore, the variations ($N\delta f, f\delta N$) exhibit \textit{thermodynamic conjugacy} similar to that in the corresponding variations $(T\delta s, s\delta T)$. 

\end{itemize}

\subsection{The Avogadro Number of the Spacetime and the Spacetime Evolution}\label{sec:avogadro}

A crucial relation in the study of, say, gases is the equipartition law $E=(1/2)Nk_BT$, which should be more appropriately written as:
\begin{equation}
Nk_B=\frac{E}{(1/2)T} 
\end{equation} 
Here, both the variables in the right-hand side, $E$ and $T$, have valid interpretations in the continuum, thermodynamic limit, but the $N$ in the left-hand 
side has no meaning in the same limit. The $N$ actually counts the microscopic degrees of freedom or, more figuratively, the number of atoms, the very existence of which is not recognized in thermodynamics! An equation like this directly relates the macroscopic and microscopic descriptions. Can we obtain a similar relation for spacetime? Can we count the number of atoms of spacetime? 

It turns out that indeed we can \cite{A17,A18}, and the current $J^a[\xi]$, where $\xi^a$ is the time evolution vector related to the (1 + 3) foliation, shows the way. Consider a section of a spacelike surface $\mathcal{V}$ with boundary $\partial\mathcal{ V}$ corresponding to $N=$ constant. In any static spacetime, one can show that the gravitating (Komar) energy $E_{\rm Komar}$ of this bulk is equal to the equipartition heat energy of the surface we encountered earlier in \eq{equiE}:
\begin{equation}
 E_{\rm Komar}\equiv\int d^{3}x\sqrt{h} \ 2N\bar{T}_{ab}u^{a}u^{b}=\int \frac{\sqrt{\gamma} \, d^2x}{L_P^2} \left( \frac{1}{2}k_B T_{\rm loc}\right) 
=\frac{1}{2} N_{\rm sur} (k_BT_{\rm avg})
\label{hequi1}
\end{equation} 
where $\bar{T}_{ab}\equiv {T}_{ab}-(1/2)g_{ab}T$.
Therefore, there is a correspondence between the bulk and boundary energies, as well as equipartition, which I will call holographic equipartition.

It gets better. When we consider the \textit{most general} spacetime (rather than static spacetimes), we would expect the above relation to break down and the difference between the two energies to drive the evolution of the spacetime. This is precisely what happens. One can associate with the bulk energy $E_{\rm Komar}$ the number $N_{\rm bulk}$, defined as the number of degrees of freedom in a bulk volume \textit{if} the (Komar) energy $E_{\rm Komar}$ contained in the bulk is at equipartition at the temperature $T_{\rm avg}$. That is:
\begin{align}\label{Papper06_NewFin01}
 N_{\rm bulk}\equiv\frac{\epsilon}{(1/2)T_{\rm avg}}\int d^{3}x\sqrt{h} \ 2N\bar{T}_{ab}u^{a}u^{b}=\frac{|E_{\rm Komar}|}{(1/2)T_{\rm avg}}
\end{align}
where $\epsilon=\pm$ is chosen so as to keep $N_{\rm bulk}$ positive, even if $E_{\rm Komar}$ turns negative.
We do \textit{not}, of course, assume that the equipartition is actually realized; this is just a dimensionless measure of the Komar energy in terms of the average boundary temperature. 

One can then show \cite{A19} that the time evolution of spacetime geometry in a bulk region, bounded by the $N=\textrm{constant}$ surface, is driven by the suitably-defined bulk and boundary degrees of freedom. Specifically:
\begin{align}\label{Paper06_NewFin02}
\frac{1}{8\pi}\int d^{3}x\sqrt{h}u_a g^{ij}\pounds_\xi N^a_{ij} =\frac{\epsilon}{2}T_{\rm avg}\left(N_{\rm sur}-N_{\rm bulk}\right)
\end{align}
with $\xi_a=Nu_a$ being the time evolution vector, where $u_a$ is the velocity of the observers moving normal to the foliation.\footnote{The Lie variation term in \eq{Paper06_NewFin02} is closely connected with the canonical structure \cite{A19} of general relativity in the conventional approach, through the relation
$ \sqrt{h}u_{a}g^{ij}\pounds _{\xi}N^{a}_{ij}=-h_{ab}\pounds _{\xi}p^{ab}$,
where $p^{ab}=\sqrt{h}(Kh^{ab}-K^{ab})$ is the momentum conjugate to $h_{ab}$ in the standard approach.} This result shows that \textit{it is the difference between the surface and the bulk degrees of freedom that drives the time evolution of the spacetime!}
(A very similar result holds \cite{SCTPnull} for a null surface, as well, in terms of corresponding variables.)

A simple, but remarkable corollary is that in all static \cite{A17,A18} spacetimes, we have holographic equipartition, leading to the equality of the number of degrees of freedom in the bulk and boundary:
\begin{equation}
 N_{\rm sur}=N_{\rm bulk}; \qquad (\mathrm {holographic\; equipartition})
 \label{nsureqn}
\end{equation} 
which, of course, is a restatement of \eq{hequi1}.

\subsection{The Fluid Mechanics of the Null Surfaces}

From the $J^a[v]$, one can define another 
vector field $P^a[v]$, which can be thought of as the gravitational momentum attributed to spacetime \cite{A36}, by an observer with velocity $v^a$. It is defined as:
\begin{equation}
 P^a[v] \equiv 2 G^a_b v^b - J^a [v] = -Rv^{a}-g^{ij}\pounds _{v}N^{a}_{ij}
 \label{deffourvel}
\end{equation}
The physical meaning of $P^a[v]$ arises from the following fact: the conservation of total momentum $(P^a + M^a)$ for all observers will lead to \cite{A36} the field equations of general relativity; the introduction of $P^a(v)$ restores the conservation of momentum in the presence of gravity!
When evaluated for the time evolution vector in for the Gaussian null coordinates (GNC)\footnote{The GNC system generalizes the notion of the local Rindler frame associated with an arbitrary null surface; see~\cite{A41,A42,A43} for more details. We define
the time development vector as $\xi^a=Nu^a$, where $u^a$ is the velocity of observers at rest in GNC. 
One can show that $\xi^a$ will reduce to the timelike Killing vector corresponding to the Rindler time coordinate if we rewrite the standard Rindler metric in the GNC form. Therefore, $\xi^a$ is a natural generalization of the time evolution vector, corresponding to the local Rindler-like observers in the GNC, though, of course, in the general case, $\xi^a$ will not be a Killing vector in a general spacetime.} associated with a given null surface, $P^a[\xi]$ reveals its thermodynamic significance in two contexts. 
First, we can show that the variational principle used to obtain the field equations has a simple interpretation \cite{SCTPnull} in terms of $P^a[\xi]$ in GNC. 
Second, the projection of $P^a[\xi]$ along $\ell_a,k_a$ and $q_{ab}$ associated with a null surface leads to three sets of equations, all of which have a direct thermodynamic interpretation.

Let us start with the variational principle, which was based on \eq{Qtot}. The $Q_{tot}$ has a simple expression in terms of the total momentum flux through the null surfaces. We can show \cite{SCTPnull} that: 
\begin{align}\label{Paper06_New_21}
{Q}_{\rm tot}=-\int d^{2}x d\lambda \sqrt{\gamma}\, \ell_a \,P^{a}_{\rm tot}(\xi)
=-\int d^{2}x d\lambda \sqrt{\gamma}\, \ell_a \,\left[P^{a}(\xi)+M^{a}(\xi)\right]
\end{align}
where the expressions in the integrand can be thought of as the limit of $\xi_a P_{\rm tot}^a(\xi)$ as we approach the null surface, and we have ignored the end point contributions. Clearly, the total energy density associated with the total momentum $P^a_{\rm tot}$, by the local Rindler observer, is what contributes to the $Q_{\rm tot}$ of the null surface. The variational principle thus has a clear physical significance \textit{even off-shell}, unlike, for example, the action principle for gravity in the conventional approach. 

The second feature about the gravitational momentum $P^a(\xi)$ is somewhat more technical, and hence, I will only mention its physical content. 
Given the thermodynamic properties of the null surfaces, one would expect the flow of gravitational momentum vis-\`{a}-vis any null surface to be of primary importance. To explore this, we construct the GNC associated with the given null surface and 
the $P^a(\xi)$ using the corresponding time evolution vector. The natural basis vectors associated with the null surface are given by the set of vectors $(\ell ^{a},k^{a},e^{a}_{A})$ where $e^{a}_{A}$ spans the two transverse directions. The gravitational momentum can be decomposed using this basis as: $P^{a}=A\ell ^{a}+Bk^{a}+C^{A}e^{a}_{A}$; and the components $A,B$ and $C^{A}$ can be recovered from the projections of $P^a$ given by $A=-P^{a}(\xi)k_{a}$, $B=-P^{a}(\xi)\ell _{a}$ and $C^{A}=P^{a}(\xi)e^{A}_{a}$. Therefore, the following combinations, $q^{a}_{b}P^{b}(\xi), k_{a}P^{a}(\xi)$ and $\ell_{a}P^{a}(\xi)$, will give the complete information about the flow of gravitational momentum vis-\`{a}-vis the given null surface. Each of them leads to an interesting thermodynamic interpretation. However, since the calculations are somewhat involved, I will skip the algebraic details (which can be found in \cite{SCTPnull}) and summarize the results:

\begin{itemize}

\item The component $q^{b}_{a}P^{a}(\xi)$ allows us to rewrite the relevant component of the field equations in a form identical to the Navier--Stokes equation for fluid dynamics \cite{A24,A25} (for a variable that can be interpreted as drift velocity on the horizons). This is probably the most direct link between the field equations and the fluid mechanics of atoms of spacetime. This result generalizes the corresponding result, known previously for black hole spacetimes \cite{A26,A27}, to \textit{any} null surface in \textit{any} spacetime. 

\item The projection $k_{a}P^{a}(\xi)$, evaluated on an arbitrary null surface, can be \cite{A33} rewritten in the form: $TdS=dE+PdV$, \textit{i.e.}, as a thermodynamic identity. Here, all of the variables have the conventional meanings, and differentials are interpreted as changes in the relevant variables when we make an infinitesimal virtual displacement of the null surface in the direction of $k^a$. This generalizes the corresponding results, previously known for spacetimes with some symmetry \mbox{(see, e.g., \cite{A30,A31,A32,A28,A29})}, and the null surface in question is a horizon. This result also allows us to associate the notion of energy with an arbitrary null surface \cite{A44,A45}.

\item Finally, the component $\ell _{a}P^{a}(\xi)$ gives \cite{SCTPnull} the evolution of null surface, in terms of its heating rate involving both $ds/d\lambda$ and $dT/d\lambda$, where $s$ is the entropy density, $T$ is the temperature associated with the null surface and $\lambda$ is the parameter along the null generator $\ell _{a}$. 

\end{itemize}

\section{A Closer Look at the Atoms of Spacetime}\label{sec:gravheatden}

The results described so far suggest that the dynamics of spacetime is the thermodynamic limit of the statistical mechanics of microscopic degrees of freedom, which we shall call the atoms of spacetime. Our~next task is to obtain the heat density $\mathcal{H}_g$, used in the variational principle based on \eq{Qtot}, from a reasonable model for microscopic degrees of freedom. Given the enormous conceptual complications in any such attempt, we will approach the problem in a step-by-step manner, proceeding by analogy with more familiar situations.

Let us start by recalling certain features in the description of a normal fluid made of atoms. The macroscopic, thermodynamic description ignores the existence of discrete structures and describes the fluid as a continuum using variables, like density $\rho(t, \mathbf{x})$, pressure $p(t, \mathbf{x})$, mean velocity $\mathbf{V}(t, \mathbf{x})$, \textit{etc}. The price we pay for ignoring the discrete structures is that we need to add certain variables (like temperature) purely phenomenologically, say through the equation of state $P=P(\rho,T)$, for this description to work properly. The next layer of description for a fluid, used in physical kinetics, is in terms of the distribution function $f(t, \mathbf{x},\mathbf{v})$. 
(In a relativistic case, we will use $f(x^i,p_j)$ with $p^ip_i=m^2$, which reduces again to $f(t, \mathbf{x},\mathbf{v})$ with a suitable Jacobian).
This description recognizes the fact that the fluid \textit{is} made of atoms. However, it works at a scale sufficiently large compared to the inter-atomic distance, so that we can interpret $dN = f(t, \mathbf{x},\mathbf{v})d^3\mathbf{x} d^3\mathbf{v}$ as the number of atoms in a phase volume $d^3\mathbf{x} d^3\mathbf{v}$. The key assumption is that we can introduce a volume element $d^3\mathbf{x}$, which is sufficiently small to be treated as `practically' infinitesimal and yet large enough to contain a sufficiently large number of atoms of the fluid.

The main difference between the descriptions in these two layers (thermodynamics \textit{vs}. physical kinetics) lies in the fact that the latter allows us to handle the dispersion in the microscopic variables. For~example, $f(t, \mathbf{x},\mathbf{v})$ tells us that, at a given location $\mathbf{x}$, there could be several atoms moving in different directions with different speeds, thereby leading to velocity dispersion. One could therefore compute \textit{both} the mean velocity \textit{and} the velocity dispersion using $f(t, \mathbf{x},\mathbf{v})$ by:
\begin{equation}
\mathbf{V}(t,\mathbf{x})\equiv \int \mathbf{v} f(t, \mathbf{x},\mathbf{v}) d^3\mathbf{v};\qquad
 \sigma_v^2(t,\mathbf{x})\equiv \int (\mathbf{v}-\mathbf{V})^2 f(t, \mathbf{x},\mathbf{v}) d^3\mathbf{v}
\end{equation}
and relate $\sigma_v^2$ to the temperature by, say, $k_BT\propto \sigma_v^2$. In contrast, the thermodynamic description only has the notion of the mean velocity $\mathbf{V}(t,\mathbf{x})$ of the fluid at an event, but not that of any velocity dispersion, since no discrete structure is recognized. As a result, we have to introduce the temperature (and other variables) in an \textit{ad hoc} manner in such a description. Clearly, the description in terms of a distribution function, recognizing the existence of atoms with different velocities at a given point, is one level closer to reality and is the first step in incorporating the discreteness at the microscopic level.

What we seek is a similar description, for the atoms of spacetime, so that we are led to the correct form of the heat density. Working from the macroscopic scales, we know that the auxiliary vector field $n_a$ plays a crucial role. 
However, the discussion in Section~\ref{sec:elegantgravdyn} shows that one can obtain the field equations with \textit{any} null vector $n_a$. In the macroscopic limit, if we identify $n_a$ with $\ell_a$, corresponding to a null congruence, then $T_{ab}\ell^a\ell^b$ has a thermodynamic interpretation. This does not immediately suggest a
unique microscopic origin for this vector field $n_a$. There are two natural interpretations one could explore.

The first one is to think of $n_a$ as representing something analogous to the velocity $\mathbf{v}$ of the atoms that appear in the distribution function. The fact that $n_a$ is null implies that the atoms of spacetime have no mass scale associated with them, which makes sense. However, in that case, one would have expected the kinetic energy contribution to the gravitational heat density to be of the form:
\begin{equation}
\mathcal{K}_{g}=\frac{1}{2}M_{ab}n^an^b
\label{hk} 
\end{equation} 
rather than a quadratic in the \textit{derivatives} of $n_a$. 

The second possibility is to think of $n_a(x)$ as analogous to the mean velocity field $\mathbf{V}(t,\mathbf{x})$, which appears at the thermodynamic description. Then, one can relate a quadratic term in $\nabla_an_b$ to some kind of viscous heat generation
(as indicated by the correspondence with the Navier--Stokes equations~\cite{A24,A25} mentioned earlier)
contributing to the heat density. In the description of normal fluids, these two are \textit{completely} different constructs. However, in the description of spacetime, we have only one kind of vector field, $n_a$, and it should somehow play roles analogous to both $\mathbf{v}$ and $\mathbf{V}(t,\mathbf{x})$ simultaneously!
Then, both of the descriptions will be valid, and we will have a natural interpretation of the heat density, both from microscopic and macroscopic scales. 
Mathematically, this requires that we should be able to express the heat density $\mathcal{H}_g$ in \eq{hgrav} in an equivalent form 
as a quadratic in $n_a$ (like \eq{hk}) without any derivatives. \textit{This is a very nontrivial constraint}; but again, everything works out fine! Let me explain how this comes about in some detail.

To do this, let us begin by asking the question:
How come the variation of a quadratic in $\nabla_an_b$, in \eq{hgrav}, did not lead to second derivatives of $n_a$ in the Euler--Lagrange equations? Algebraically, this is due to the occurrence of the commutator of covariant derivatives $[\nabla_i, \nabla_j] n_k$, which, as we know, is linear in $n_m$ and does not contain any second derivatives. There is, however, a nicer way to see this result \cite{A19}, which is based on the following identity: 
\begin{equation}
2P^{ab}_{cd}\nabla_an^c\nabla_bn^d= R_{ab}n^an^b + \frac{1}{\sqrt{\gamma}} \frac{d}{d\lambda} (\sqrt{\gamma}\Theta)
\label{prsep}
\end{equation}
where $n^i$ is the affinely parameterized congruence with $\lambda$ being the affine parameter and $\Theta = \nabla_in^i = (d/d\lambda)(\ln \sqrt{\gamma})$. Therefore, $\mathcal{H}_g$ and $R_{ab}n^an^b$ differ by a total derivative term that does not contribute to the variation, and we can write: 
\begin{equation}
 Q_{\rm tot} = \int \sqrt{\gamma}\, d^2x d\lambda\left[ - \frac{R_{ab}}{8\pi L_P^2} + T_{ab} \right] n^an^b - \frac{1}{8\pi L_P^2} \int d^2 x \, \sqrt{\gamma}\, \Theta \Bigg|^{\lambda_2}_{\lambda_1} 
\label{var1}
\end{equation} 
Ignoring the second term, since it contributes only at the end points $\lambda = (\lambda_1,\lambda_2)$, our variational principle reduces to working with $(-R_{ab}/8\pi L_P^2 + T_{ab})n^an^b$. Imposing the $n^2 =0$ condition by a Lagrange multiplier
and varying this expression with respect to $n^a$ will lead to $R^a_b = (8\pi L_P^2) T^a_b + f(x) \delta^a_b$. Taking the divergence and using the Bianchi identities, as well as $\nabla_a T^a_b =0$, we find that \mbox{$f(x) = (1/2) R \ + $} constant, thereby leading to Einstein's equations with the \cc\ appearing as an integration constant. 

\eq{prsep} also shows that $\mathcal{H}_g$ reduces to a total divergence term in flat spacetime (expressed in, say, the Rindler coordinates) and isolates the contribution due to spacetime curvature, which is contained~in:
\begin{equation}
\mathcal{K}_{g} \equiv - \frac{1}{8\pi L_P^2} R_{ab} n^an^b 
\label{hk1}               
\end{equation} 
Everything would have worked out fine even if we had used an expression for the gravitational heat density\footnote{A conceptually unsatisfactory feature of the standard approach to dynamics is that it equates a purely geometrical object $G^a_b$ to a matter variable $T^a_b$. It is unclear what is common to the two sides of this equation. Our approach shows clearly what is common to both sides of Einstein's equations, if we write it as $ (8\pi L_P^2)^{-1}R_{ab}\ell^a\ell^b=T_{ab}\ell^a\ell^b$. They both represent the heat densities, of spacetime and matter! Moreover, all of these results generalize to \LL\ models with $R^a_b$ replaced by $E^a_b$, \textit{etc.}} given by  \eq{hk1}. 

The result in \eq{hk1} has exactly the same structure seen in \eq{hk}, which is what we wanted. Therefore, we could have thought of our $n_a$ as analogous to: (i) the macroscopic, mean velocity field 
$\mathbf{V}(t,\mathbf{x})$ and interpreted $\mathcal{H}_g$ in \eq{hgrav} as the heat density arising from something analogous to viscous dissipation; or (ii) the microscopic velocity field 
$\mathbf{v}$, which can be interpreted as analogous to the velocity of the atoms themselves. It is very gratifying that the same heat density allows both of the descriptions. The corresponding heating rate, made dimensionless for future convenience, is given by:
\begin{equation}
\frac{d(Q_{g}/E_P)}{d(\lambda/L_P)}= L_P^2\frac{dQ_{g}}{d\lambda}=L_P^2\int\sqrt{\gamma}d^2x \ \mathcal{K}_{g}=
-\int\frac{\sqrt{\gamma}d^2x}{L_P^2}\left(\frac{L_P^2}{8\pi } R_{ab} n^an^b\right)
\label{qrate}
\end{equation} 
In fact, one can also work with a variational principle based on $(dQ_{g}/d\lambda)$ (rather than $Q_{g}$), if we use this expression in \eq{var1}. Therefore, the variational principle can be thought of as an extremum condition on the heating rate.

It is possible to make some more progress with the expression in \eq{hk1} by recognizing that one could limit oneself to affinely parameterized null vectors $n^a =\nabla_a \sigma$, which are pure gradients. In that case, the gravitational heat density in \eq{hk1} takes the form:
\begin{equation}
\mathcal{K}_{g} \equiv - \frac{1}{8\pi L_P^2} R^{ab} \nabla_a\sigma \nabla_b\sigma 
\label{hk2}               
\end{equation} 
If we use this expression in \eq{Qtot} and vary $\nabla_a\sigma$, imposing the constraint that $\nabla_a\sigma$ is null, 
we will again get the correct field equations. As we mentioned earlier on, we really have no idea what is the extra degree of freedom $q_A$ on which our extremum principle will depend, when we approach it from the microscopic scales; 
an $n_a$ of the form $\nabla_a\sigma$ is adequate.

Therefore, our task now reduces to coming up with a microscopic model, which will have the following~features:

\begin{itemize}

\item The key new ingredient in our approach is the introduction of a vector field $n_a = \nabla_a\sigma$ into a variational principle. It is not \textit{a priori} clear how the auxiliary variable, like $\sigma$ or $n_a$, arises from a microscopic description
and why we need to vary it in an extremum principle. The microscopic description should lead to 
the vector field $n_a = \nabla_a \sigma$, as well as $\sigma$ itself. This is probably the most important task. 

\item There should be a fundamental reason why null vectors, closely associated with local Rindler horizons, play such an important role. This should emerge from the microscopic description. 

 \item Finally, we need to obtain the \textit{explicit} form of the heat density in \eq{hk2} in a natural manner from the microscopic description. 

\end{itemize}
These might appear to be fairly formidable tasks, but I will show that it is possible to come up with a microscopic description that satisfies all of these criteria!

It turns out that $\sigma$, as well as the combination $R^{ab} \nabla_a\sigma \nabla_b\sigma $ have a very natural interpretation, which I will now describe.
To do this, I want to introduce an alternate way of describing the standard Riemannian geometry using what is known \cite{E5,E8,E9,E10} as Synge's world function $\sigma^2(x,x')$, instead of the metric tensor $g_{ab}(x)$. The world function $\sigma^2(x,x')$ is defined as the geodesic interval between any two events $x$ and $x'$, which are sufficiently close so that a unique geodesic exists. Since the knowledge of all geodesic distances (locally) is equivalent to the knowledge of the metric, anything one can do with the metric tensor can be done using $\sigma^2(x,x')$. The information contained in the ten functions $g_{ab}(x)$, which depends on the choice of the coordinate system, is more efficiently encoded in the single biscalar $\sigma^2(x,x')$. (Of course, one could summarize the information of ten functions in a single function only because $\sigma^2$ is nonlocal and depends on \textit{two} events $x$ and $x'$). Mathematically, this arises from the expansion: 
\begin{equation}
 \frac{1}{2} \nabla_a \nabla_b \sigma^2 = g_{ab} - \frac{\lambda^2}{3} \mathcal{E}_{ab} +\frac{\lambda^2}{12} n^i\nabla_i \mathcal{E}_{ab} + \mathcal{O} (\lambda^4)
\label{sigexp}
\end{equation} 
where $\lambda$ is the affine distance along the geodesic connecting $x$ and $x'$, $\mathcal{E}_{ab} \equiv R_{akbj} n^kn^j$ and: 
\begin{equation}
 n_a = \frac{1}{2\sqrt{ |\sigma^2|}}\, \nabla_a \sigma^2 = \nabla_a \sigma
\end{equation} 
(The second equality follows from the fact that $\sigma$ satisfies the Hamilton--Jacobi equation leading to $g^{ab}\nabla_a \sigma^2\nabla_b \sigma^2 = 4\sigma^2$; we will assume $\sigma^2 >0$ for simplicity when it will not cause any problems.)
 \eq{sigexp} shows that the coincidence limit ($x\to x'$) of $(1/2) \nabla_a \nabla'_b\sigma^2$ gives the metric tensor $g_{ab}$. Given the geodesic distance $\sigma^2(x,x')$, we can obtain $g_{ab}$ at any event and, hence, can calculate any other geometrical quantity. Therefore, all of gravitational dynamics can be done, in principle, with $\sigma^2(x,x')$ instead of with the metric.

The expansion in \eq{sigexp} shows that the second order term contains the combination $\mathcal{E}_{ab}$, the trace of which is given by:
\begin{equation} 
\mathcal{E} = R^{ab}n_an_b = R^{ab} \nabla_a \sigma \nabla_b \sigma 
\end{equation} 
This has an algebraic structure identical to the heat density in \eq{hk2}! This suggests that if we work with $\sigma^2(x,x')$ (rather than with the metric), then some natural variables in the microscopic theory could possibly be related to the heat density in \eq{hk2}. 

Let me illustrate how $\mathcal{E}$ occurs in several geometrical variables in a natural fashion \cite{D8}. 
To do this, we will switch from the Lorentzian spacetime to Euclidean spacetime around an event $P'$, so that:
(i) $\sigma^2 (P',P)$ treated as a function of $P$ (with fixed $P'$) is positive.
(ii) The local Rindler horizon gets mapped to the Euclidean origin, which we take to be $P'$.
(iii) The coincidence limit of $P\to P'$, approaching the origin, corresponds to approaching the local Rindler horizon in the original spacetime. 
(The coincidence limit $\sigma^2 \to 0$ corresponds to all of the events $P$ in the original spacetime connected to the origin $P'$ by a null ray.)

In the Euclidean spacetime, it is convenient to introduce the notion of an equi-geodesic surface that corresponds to all events at the same geodesic distance from the origin \cite{D1,D4,D5,D6}. To describe such a surface, it is convenient to work with a natural coordinate system $(\sigma, \theta_1, \theta_2, \theta_3)$ where $\sigma$ (the geodesic distance from the origin) is the ``radial'' coordinates and $\theta_\alpha$ are the angular coordinates on the 
equi-geodesic surfaces corresponding to $\sigma =$ constant \cite{D7}. 
The metric can then be reduced to the form: 
\begin{equation}
 ds^2_E = d\sigma^2 + h_{\alpha\beta} dx^\alpha dx^\beta 
\label{sync}
\end{equation} 
where $h_{\alpha\beta} $ is the induced metric on the equi-geodesic surface with $\sigma =$ constant.\footnote{This is the analogue of the synchronous frame in Minkowski spacetime, with $x^\alpha$ chosen to be angular coordinates.}
The most primitive quantities one can introduce in such a spacetime are the volume element $\sqrt{g}\, d^4x$ and the area element of the equi-geodesic surface, $\sqrt{h}\, d^3x$. For the metric in \eq{sync}, we, of course, have $\sqrt{g} = \sqrt{h}$, and hence, both the volume and area measures are identical. It is possible to show \cite{D8} that in the limit of $\sigma \to 0$, this measure is given by: 
\begin{align}
\sqrt{h}= \sqrt{g}=\sigma^3\left(1-\frac{1}{6}\mathcal{E}\sigma ^{2}\right)\sqrt{h_\Omega}
\label{gh}
\end{align}
where $\sqrt{h_\Omega}$ arises from the standard metric determinant of the angular part of a unit sphere. This is the simplest example of the appearance of $\mathcal{E}$ in a primitive geometrical variable. It gives the correction to the area of (or the volume enclosed by) an equi-geodesic surface. This is a very standard result in differential geometry and is often mentioned as a measure of curvature around any event. 

It seems natural to assume that the number of atoms of spacetime (\textit{i.e.}, the microscopic degrees of freedom, contributing to the heat density) at $P$ should be proportional to either the area or volume ``associated with'' the event $P$. This is because we would expect the number of atoms of spacetime to scale with either area or volume. (Based on the earlier result $N_{sur}=A/L_P^2 = N_{\rm bulk}$ in equipartition, we would expect 
a scaling with $\sqrt{h}$, which is the ``area'' element of $\sigma=$ constant surface,
but it is important to \textit{derive} this and understand why volume scaling does not arise in the microscopic description). To give precise meaning to the phrase, area or volume ``associated with'' the event $P$, we can proceed as follows: (i) we construct an equi-geodesic surface $S$ centered on $P$ with ``radius'' $\sigma$; (ii) we compute the volume enclosed by $S$ and the surface area of $S$; and (iii) we take the limit of $\sigma\to0$ to determine the area or volume associated with $P$.
However, as we can see from \eq{gh}, these measures identically vanish in the limit of $P\to P'$, which corresponds to $\sigma \to 0$. Therefore, while the required combination $\mathcal{E} = R^{ab}\nabla_a\sigma \nabla_b\sigma$ does exist in the volume and area measures, it does not contribute in the appropriate limit. 

A little thought shows that this is certainly to be expected. As we saw from the macroscopic approach, the entropy requires a quantum of area $L_P^2$ for its proper description. Classical differential geometry, which is what we have used so far, knows nothing about a quantum of area and, hence, cannot give us the correct heat density. To obtain the heat density from the above considerations, we need to ask how the geodesic interval gets modified in a quantum description of spacetime and whether such a modified description will have a $\sqrt{h}$ (or $\sqrt{g}$) leading to the correct heat density. The last miracle I will describe is how this comes about. 

\section{The Renormalized Spacetime}\label{sec:kineticsast}

The essential idea was to recognize that a primary effect of quantum gravity will be to introduce into the spacetime a zero-point length \cite{D2a,D2b,D2c,D2d,D2e,D2f}, by modifying the geodesic interval $\sigma^2(x,x')$ between any two events $x$ and $x'$ (in a Euclidean spacetime) to a form like $\sigma^2 \to \sigma^2 + L_0^2$ where $L_0$ is a length scale of the order of the Planck length.
More generally, such a modification can take the form of $\sigma^2 \to S(\sigma^2)$, where the function $S(\sigma^2)$ satisfies the constraint $S(0) = L_0^2$ with $S'(0)$ finite. (Our results are happily insensitive to the explicit functional form of such $S(\sigma^2)$; so, for the sake of explicit illustration, we will use $S(\sigma^2) = \sigma^2 + L_0^2$.) The theoretical evidence for the existence of such a zero point length is described in several previous works \cite{D2a,D2b,D2c,D2d,D2e,D2f} and will not be repeated here. While we may not know how quantum gravity modifies the classical metric, we do have an indirect handle on it if quantum gravity introduces a zero point length to the spacetime in the manner described above.

Since the original $\sigma^2$ can be obtained from the original metric $g_{ab}$ (and \textit{vice versa}), it will be nice if we can obtain the quantum gravity-corrected geodesic interval $S(\sigma^2)$ from a corresponding quantum gravity-corrected metric \cite{D1}, which we will call the q-metric $q_{ab}$. Obviously, no such local, non-singular $q_{ab}$ can exist because, for any such $q_{ab}$, the resulting geodesic interval will vanish in the coincidence limit, almost by definition. Therefore, we expect $q_{ab}(x,x')$ to be a bitensor, which is singular at all events in the coincidence~limit.
One can now determine \cite{D4,D5} the form of such a $q_{ab}(x,x')$ for a given $g_{ab}(x)$ by using two~conditions: 
(i) It should lead to a geodesic interval $S(\sigma^2)$ with a zero point length and;
(ii) The Green function describing small metric perturbations should have a non-singular coincidence limit. It can be shown \cite{D5} that these conditions determine $q_{ab}$ uniquely in terms of $g_{ab}$ (and its associated geodesic interval $\sigma^2$). We get: 
\begin{align}
q_{ab}=Ah_{ab}+ B n_{a}n_{b};\qquad q^{ab}=\frac{1}{A}h^{ab}+\frac{1}{B}n^{a}n^{b}
\label{qab}
\end{align}
where $D$ is the dimension of spacetime, $D_k$ is a shorthand for $D-k$ and:
\begin{align}
B=\frac{\sigma ^{2}}{\sigma ^{2}+L_{0}^{2}};\qquad A=\left(\frac{\Delta}{\Delta _{S}}\right)^{2/D_{1}}\frac{\sigma ^{2}+L_{0}^{2}}{\sigma ^{2}};\qquad n_a=\nabla_a\sigma
\label{defns}
\end{align}
and $\Delta$ is the Van Vleck determinant related to the geodesic interval $\sigma^2 $ by:
\begin{align}
\Delta (x,x')=\frac{1}{2}\frac{1}{\sqrt{g(x)g(x')}}\textrm{det}\left\lbrace \nabla _{a}^{x}\nabla _{b}^{x'}\sigma ^{2}(x,x') \right\rbrace
\end{align}
The $\Delta_S$ is the corresponding quantity computed with $\sigma^{2}$ replaced by $S(\sigma^{2})$ in the above definition.

Before proceeding further, I want to introduce the notion of a renormalized (`dressed') spacetime \cite{paperD} and interpret
$q_{ab}$ as the renormalized spacetime metric, which incorporates some of the non-perturbative effects of quantum gravity at Planck scales. While this is not essential for what follows, it provides a possible back drop for understanding the origin of $q_{ab}$.

An important effect of the interactions in quantum field theory is to replace the bare variables in a Lagrangian by physical variables, which incorporate (some) effects of the interactions. We know that, in general, \textit{such a renormalization changes not only the constants, which appear in the Lagrangian, but also the field variables.} For example, consider the usual $\lambda \phi^4$ theory of a scalar field in $D=4$, described by a Lagrangian $L(\phi_B;m_B,\lambda_B)$ in terms of the bare variables. The perturbation theory (carried up to the two-loop level) tells us that we need to renormalize not only $\lambda_B$ and $m_B$ to their physical values $\lambda$ and $m$, but also change the bare field $\phi_B$ to the physical field $\phi$ if the theory is to make sense. A similar effect arises in QED, as well, which requires field renormalization. Though these results are usually obtained in perturbation theory, the requirement of renormalization by itself is a non-perturbative feature. In the Wilsonian interpretation of the field theory, integrating out the high energy modes will lead to the renormalization of the low energy effective Lagrangian, which is a feature that transcends perturbation~theory. 

It seems, therefore, natural to assume that a similar effect will arise in the case of gravity, as well. The bare Lagrangian for gravity, $L(g_{ab}^B, G_B, \Lambda_B) \propto G_B^{-1}[R(g_{ab}^B) - 2\Lambda_B] \sqrt{-g_B}$ should be interpreted as being expressed in terms of \textit{not only} the bare coupling constants ($G_B$ and $\Lambda_B$), \textit{but also} the bare metric tensor $g_{ab}^B$. We would then expect quantum gravitational processes at the Planck scale to replace 
($g_{ab},G_B,\Lambda_B$) by their renormalized, physical, counterparts ($g^R_{ab}, G, \Lambda$). We can then compute all other renormalized geometrical variables (e.g., the curvature tensor) by using the $g^R_{ab}$ in the place of $g^B_{ab}$ in the relevant expressions. This procedure is necessarily approximate, compared to a fully rigorous non-perturbative quantum gravitational approach, which we do not have, but will surely capture some of the effects at the intermediate (``mesoscopic'') scales between the Planck scale and the long wavelength limit at which the classical metric is adequate. 
Of course, we cannot use perturbation techniques to directly compute $g^R_{ab}$ for a given classical geometry described by a $g_{ab}$, and we would expect $g^R_{ab}$ to be non-local and singular at any given event. (We will drop the superscript $B$ in $g_{ab}^B$ hereafter.) However, since the same quantum gravity effects that replace $g_{ab}$ by $q_{ab}$ are expected to replace $\sigma^2$ by $S(\sigma^2)$, we can identify $g^R_{ab}=q_{ab}$ in \eq{qab}. Therefore, we have an indirect way of determining the renormalized spacetime $g^R_{ab}=q_{ab}$ by this procedure.

Let us get back to $q_{ab}$. As shown in previous work \cite{D1,D6}, the q-metric has several interesting properties, which I will now list: 

(1) The $q_{ab}(x,x')$, unlike $g_{ab}(x)$, is a bitensor depending on \textit{two} events through $\sigma^2(x,x')$. As we said before, this non-locality is essential if spacetime has to acquire a zero-point length. Any local, nonsingular metric will lead to a $\sigma^2(x,x')$, which vanishes in the limit of $x\to x'$.

(2) The $q_{ab}$ reduces to the background metric $g_{ab}$ in the limit of $L_0^2 \to 0$, as it should. In the opposite limit of $(\sigma^2 / L_0^2) \to 0$, the $q_{ab}$ is singular, which is again natural if we interpret $q_{ab}$ as the metric of the renormalized spacetime; it is not expected to be well defined at any localized event and will require some kind of smearing over Planck scales.

(3) When $g_{ab}=\delta_{ab}$, the $q_{ab}$ is also locally flat in the sense that there exists a coordinate transformation, which will reduce $q_{ab} dx^a dx^b$ to $\eta_{ab} dx^a dx^b$ in the synchronous frame. (This is, however, rather subtle because the coordinate transformation removes a region of size $L_P$ from the spacetime around \textit{all} events.)

(4) Let $\Phi[g_{ab}(x)]$ be some scalar or scalar density (like, for example, the Ricci scalar $R[g_{ab}(x)]$) constructed from the background metric and its derivatives. 
We can compute the corresponding 
(bi)scalar $\Phi[q_{ab}(x,x');L_0^2]$ for the renormalized spacetime by replacing $g_{ab}$ by $q_{ab}$ in $\Phi[g_{ab}(x)]$ and evaluating all of the derivatives at $x$ keeping $x'$ fixed. The renormalized value of $\Phi[q_{ab}(x,x');L_0^2]$ is obtained by taking the limit $x\to x'$ in this expression keeping $L_0^2$ non-zero. Several useful scalars like $R$, $K$, \textit{etc.}, remain finite~\cite{D1,D5,D6}
and local in this limit, even though the q-metric itself is singular when $x\to x'$ with non-zero $L_0^2$. The algebraic reason for this result~\cite{D1} is that the following two limits do not commute:
\begin{equation}
\lim_{L_0^2\to 0}\, \lim_{x\to x'} \Phi[q_{ab}(x,x');L_0^2]\neq \lim_{x\to x'}\,\lim_{L_0^2\to 0} \Phi[q_{ab}(x,x');L_0^2]
\end{equation} 
All of the computations involving the $q_{ab}$ are most easily performed \cite{D7} by choosing a synchronous frame for the background metric, given in \eq{sync}, which can always be done in a local region.

\section{A Point Has Zero Volume but Finite Area!}\label{sec:eventarea}

We will now re-evaluate the area element of an equi-geodesic surface and the volume element for the region enclosed by it 
using the renormalized q-metric. This will involve $\sqrt{q} \ d^4x$ and $\sqrt{h}\, d^3 x$ as the respective integration measures, where $h$ now stands for the determinant of the induced metric on the equi-geodesic surface, corresponding to $q_{ab}$. (For the q-metric in \eq{qab}, calculated for the $g_{ab}$ in \eq{sync}, these two measures will not be equal, because $q_{00} \neq 1$.)
If our ideas are correct, $\sqrt{h}$ should lead to the correct density of the atoms of spacetime in the coincidence limit. Further, there must be a valid mathematical reason to prefer the area element $\sqrt{h}$ over the volume element $\sqrt{q}$. I will show that these hopes are indeed realized! 

It is straightforward to compute these quantities using the q-metric, and we find that (with \mbox{$S(\sigma^2)=\sigma^2+L_0^2$} chosen for illustration, though the final results \cite{D7} hold in the more general case, as well as in $D$ dimensions):
\begin{align}
\sqrt{q}=\sigma \left(\sigma ^{2}+L_{0}^{2}\right)\left[1-\frac{1}{6}\mathcal{E}\left(\sigma ^{2}+L_{0}^{2}\right)\right]\sqrt{h_\Omega}
\label{qfinal}
\end{align}
and:\footnote{
This result is
algebraically subtle. One might think that the expression in \eq{hfinal} (which is actually $\sqrt{h}=A^{3/2}\sqrt{g}$) might arise from the standard result in differential geometry, \eq{gh}, by the replacement $\sigma^2\to(\sigma^2+L_{0}^{2})$. However, note that this trick does \textit{not} work for the expression in \eq{qfinal} (which is $\sqrt{q}=\sqrt{B}A^{3/2}\sqrt{g}$) due to the $\sqrt{B}=\sigma(\sigma ^{2}+L_{0}^{2})^{-1/2}$ factor that has the limiting form $\sqrt{B}\approx\sigma/L_{0}$ when $\sigma\to0$. This is the key reason why the event has zero volume, but finite area. A possible insight into this, rather intriguing, feature is provided by the following fact:
The leading order dependence of $\sqrt{q}d\sigma\approx\sigma d\sigma$ makes the volumes scale as $\sigma^2$ (while the area measure is finite). This, in turn, is related to the fact \cite{paperD} that \textit{the effective dimension of the renormalized spacetime becomes $D=2$ close to Planck scales}, a result that has been noticed by several \mbox{people~\cite{z1,z2,z3,z4}}
in different, but specific, models of quantum gravity. Our approach seems to give this result in a \textit{model-independent} manner, which, in turn, is the root cause of the result that events have zero volume, but finite area.}
\begin{align}
\sqrt{h}=\left(\sigma ^{2}+L_{0}^{2}\right)^{3/2}\left[1-\frac{1}{6}\mathcal{E}\left(\sigma ^{2}+L_{0}^{2}\right)\right]\sqrt{h_\Omega}
\label{hfinal}
\end{align}
When $L_{0}^{2}\to0$, we recover the result in \eq{gh}, as we should. However, as explained in Item~(4) above, our interest is in the limit $\sigma^2\to0$ at finite $L_P$.
Something remarkable happens when we do this. The volume measure $\sqrt{q}$ vanishes (just as in the case of the original metric), showing that it cannot lead to anything nontrivial. The zero point length does not lead to a residual volume measure.
However, in the limit of $\sigma^2 \to 0$, we find that $\sqrt{h}$ has a non-zero limit! It is given~by:
\begin{align}
\sqrt{h}= L_{0}^{3}\left[1-\frac{1}{6}\mathcal{E}L_{0}^{2}\right]\sqrt{h_\Omega}
\label{hlimit}
\end{align}
As the title to this section indicates, the q-metric (which we interpret as representing the renormalized spacetime)
attributes to every point in the spacetime a finite area measure, but a zero volume measure! 
Since $L_0^3\sqrt{h_\Omega}$ is the volume measure of the $\sigma=L_0$ surface, the dimensionless density of the atoms of spacetime, contributing to the gravitational heat, can be taken to be:
\begin{equation}
f(x^i,n_a)\equiv \frac{\sqrt{h}}{L_0^3\sqrt{h_\Omega}} =1-\frac{1}{6}\mathcal{E}L_{0}^{2}
=1-\frac{1}{6} L_{0}^{2} R_{ab}n^an^b
\label{denast}
\end{equation} 

How can we interpret this expression for the ``number of atoms of spacetime''? 
Our intention all along has been to define the analogue of a distribution function $f(x^i,n_a)$ that gives the number of
atoms of spacetime at \textit{a given event}. We expected $f(x^i,n_a)$ to depend on an auxiliary vector field $n_a$, as well as on the location $x^i$. Just as in the usual kinetic theory, we no longer think of this location as a mathematical point, but imagine a region that contains a sufficiently large number of atoms of spacetime to make the description in terms of $f(x^i,n_a)$ valid. (To think of spacetime being filled with atoms is no stranger than thinking of matter being filled with atoms; both descriptions work at scales larger than the inter-atomic spacing, but recognize the existence of discrete structures.) The dependence on $x^i$ can have a universal part (which could exist even in the flat spacetime limit), as well as a part that depends on (what we call in macroscopic physics) the spacetime curvature. Since we want $f(x^i,n_a)$ to arise from the most basic of the geometrical properties of the space, it seems reasonable to explore areas and volumes. We know from classical differential geometry that areas and volumes of a region of size $r$ do have a flat space contribution, which is corrected by curvature-dependent terms. However, now, we want the area $(\sqrt{h}d^3x)$ and volume $(\sqrt{g} d^4x)$ measures to be defined \textit{at a point}, which will require taking the limit $r\to 0$. In a classical spacetime, both the measures vanish in this limit, as to be expected. When we consider the renormalized spacetime incorporating a zero point length, one might have naively expected \textit{both} of them to be finite at a given event. Remarkably enough, the volume measure ($\sqrt{q} d^4x$) still vanishes when the region collapses to a point, but the area measure does not.\footnote{One likes to think of the number of atoms per unit \textit{spatial} volume, rather than unit \textit{spacetime} volume, whether it is atoms of a gas or a spacetime; this is what we get from $\sqrt{h}\, d^3x$.}
Briefly stated, quantum gravity endows each event in spacetime with a finite area, but zero volume. It is this area measure that we compute to obtain a natural estimate for $f(x^i,n_a)$.

The desirable, but intriguing feature of this result is that a vector field $n_a = \nabla_a \sigma$ has survived in the final expression. At any given event (to which the coincidence limit has been taken), this vector field can point in all directions, because the geodesics emanating from that event can point in all directions. Therefore, the function $f(x^i,n_a)$ depends on the choice of the vector field $n_a$ at a given event. This is, again, very reminiscent of the distribution function $f(x^i, p^j)$ for a bunch of relativistic particles, which gives the number of particles at an event $x^i$ with momentum $p^j$. As I have emphasized earlier, the coexistence of several particles, with different momenta, at a given event is the characteristic feature of the description in physical kinetics. This assumes that one can consider a volume $d^3 \mathbf{x}$ that is small enough to be treated as infinitesimal, but large enough to contain several particles. In the same spirit, we should think of $f(x^i,n_a)$ as the number of atoms of spacetime, or less figuratively, the number of microscopic degrees of freedom, at an event $x^i$ with an extra attribute $n_a$, which is analogous to the momentum that appears in the distribution function in physical kinetics.\footnote{Incidentally, a field redefinition $g^{ab} \to g^{ab} - (L_0^2/6) R^{ab}$ in $g^{ab} \nabla_a \sigma\nabla_b \sigma$ will lead to \eq{denast}; similar field redefinitions have been used (see, e.g., \cite{tpr2}) in quantum gravity, but the connection with our approach is unclear.}

\textit{It is also easy to see how null surfaces and null vectors are singled out in this approach.} This is because the coincidence limit $P'\to P$ in the Euclidean sector (with the event $P$ taken to be the origin) corresponds to approaching the null horizon in the Minkowski sector. In all calculations, we will eventually take the limit $\sigma^2 \to 0$ in the Euclidean sector.
However, this limit, $\sigma^2 \to 0$, will translate into a null surface in the Minkowski spacetime.\footnote{The local Rindler observers who live on the hyperboloid $r^2-t^2=\sigma^2$ see the null cone $r^2-t^2=0$ as the horizon. In the Euclidean sector, the hyperboloid becomes the sphere $r^2+t_E^2=\sigma_E^2$, and approaching the Euclidean origin, $\sigma_E\to0$, translates to approaching the light cone in the Minkowski space.} The normal vector $n_i = \nabla_i\sigma$ (which occurs in the q-metric and all of the resulting constructs) will pick out the null vector, which is the normal to the null surface. More generally, $\sigma^2 (x,x') \to 0$ selects out events that are connected by a null geodesic, and hence, $n_a$ will correspond to a null vector in the Minkowski spacetime.
This is how a null vector field $n_i$ is introduced in the description from a microscopic point of view.

 It is also understandable that we should extremize the expressions with respect to this variable, which is, in some sense, the relic from quantum gravity. In fact, the extremum condition is equivalent to demanding that $Q_{g}$ should not depend on this arbitrary vector field $n_a$, which is another way of interpreting the variational principle. 

Let us complete the analysis by connecting up with the macroscopic limit. 
The contribution to the gravitational heat in any volume is obtained by integrating $f(x^i,n_j)$ over the volume. Therefore, the expression for the heating rate, in dimensionless form (corresponding to \eq{qrate}), is given by:
\begin{equation}
L_P^2\frac{dQ_{g}}{d\lambda}=\int\frac{\sqrt{\gamma}d^2x}{L_P^2}f(x^i,n_j) =
\int\frac{\sqrt{\gamma}d^2x}{L_P^2}\left[1-\frac{1}{6}L_{0}^{2}(R_{ab}n^an^b)\right]
\label{corres}
\end{equation} 
which gives the the correct expression in \eq{qrate}, with the crucial minus sign, plus a constant\footnote{If we had used, say, $\mu L_P$, rather than $L_P$ in \eq{corres} to obtain the dimensionless result here (and retained $L_P$ in \eq{qrate}), the constant term will become $\mu^{-4}$, and we will get $L_0^2=(3/4\pi)\mu^4 L_P^2$; we choose $\mu=1$ to get the unit degree of freedom as the constant term. It is also possible to add a proportionality constant in \eq{denast} which we have set to unity.} if
we set
$L_0^2=(3/4\pi)L_P^2$. Thus, the consistency of the macroscopic and microscopic descriptions also allows us to determine the value of the zero point length in terms of $L_P$ (which we know observationally from the Newtonian limit). 
\textit{Therefore, one can indeed interpret the gravitational heat density from the area measure of the renormalized~spacetime.}

While the second term in \eq{corres} gives what we want for the variational principle, the first term is important for two reasons:
\begin{itemize}
\item It tells us that there is a zero-point contribution to the degrees of freedom in spacetime, which, in dimensionless form, is just unity. Therefore, it makes sense to ascribe $A/L_P^2$ degrees of freedom to an area $A$, which is consistent with what we saw in the macroscopic description.
 \item The result tells us that a two sphere of radius $L_P$ has $4\pi L_P^2/L_P^2=4\pi$ degrees of freedom. This was the crucial input that was used in a previous work to determine the numerical value of the \cc\ for our universe. Thus, the microscopic description does allow us to determine \cite{C8,C9} the value of the \cc, which appeared as an integration~constant. 
\end{itemize}

Let me elaborate a bit on the last point, since it can provide a solution to what is usually considered the most challenging problem of theoretical physics today. 

Observations indicate that our universe is characterized by three distinct phases: (i) an inflationary phase with approximately constant density $\rho_{inf}$; (ii) a phase dominated by radiation and matter, with $\rho=\rho_{eq}[x^{-4}+x^{-3}]$, where $x(t)\equiv a(t)/a_{\rm eq}$, the $\rho_{eq}$ is a (second) constant and $a_{eq}$ is the epoch at which the matter and radiation densities were equal; and (iii) an accelerated phase of expansion at late times driven by the energy density $\rho_\Lambda$ of the cosmological constant. Values of these three constants $[\rho_{inf},\rho_{eq},\rho_\Lambda]$ will completely specify the dynamics of our universe. Standard high energy physics can, in principle, determine $\rho_{inf}$ and $\rho_{eq}$, but we need a new principle to fix the value of $\rho_\Lambda$, which is related to the integration constant that appears in our approach to field equations.

It turns out that such a universe with these three phases has a new \textit{conserved} quantity, \textit{viz}. the number $N$ of length scales, which cross the Hubble radius during any of these phases \cite{C8,C9}. Any physical principle that determines the value of $N$ during the radiation-matter dominated phase, say, will determine $\rho_\Lambda$ in terms of $[\rho_{inf},\rho_{eq}]$. The emergent paradigm tells us that the value of this conserved quantity $N$ can be fixed at the Planck scale as the degrees of freedom in a two-sphere of radius $L_P$, giving $N=4\pi L_P^2/L_P^2 = 4\pi$. This, in turn, leads to the remarkable prediction relating the three \mbox{densities~\cite{C8,C9}:}
\begin{equation}
 \rho_\Lambda \approx \frac{4}{27} \frac{\rho_{\rm inf}^{3/2}}{\rho_{\rm eq}^{1/2}} \exp (- 36\pi^2)
\label{ll6}
 \end{equation} 
From cosmological observations, we find that $\rho_{eq}^{1/4} = (0.86 \pm 0.09) \text{eV}$; if we take the range of the inflationary energy scale as $\rho_{\rm inf}^{1/4} = (1.084-1.241) \times 10^{15}$ GeV, we get $\rho_{\Lambda} L_P^4 = (1.204 - 1.500) \times 10^{-123}$, which is consistent with observations! 
This novel approach for solving the cosmological constant problem provides a unified view of cosmic evolution, connecting all three phases through \eq{ll6}; this is to be contrasted with standard cosmology in which the three phases are put together in an unrelated, \textit{ad hoc} manner.

Further, this approach to the cosmological constant problem \textit{makes a falsifiable prediction}, unlike any other approach I know of. From the observed values of $\rho_\Lambda$ and $\rho_{\rm eq}$, we can constrain the energy scale of inflation to a very narrow band, to within a factor of about five, if we consider the ambiguities in re-heating. If future observations show that inflation took place at energy scales outside the band of $(1-5)\times 10^{15}$ GeV, this model for explaining the value of \cc\ is ruled out.

\section{Discussion and Outlook}\label{sec:summary}

The paradigm described here has two logically distinct parts. The first part (Sections~\ref{sec:gravemerge}--\ref{sec:geotherm}) is mathematically rigorous and paints an alternative picture about the nature of gravity. It is based on the desire to have a strong physical principle to describe the dynamics of gravity, \textit{viz}. that the field equations should be invariant under the shift $T^a_b \to T^a_b + ({\rm constant})\ \delta^a_b$. This principle is powerful enough to rule out the metric as a dynamical variable and suggests that any variational principle that we use should depend on the matter sector through the combination $T^a_b \ell_a\ell^b$ where $\ell^a$ is a null vector. This combination is interpreted by the local Rindler observers as the heat density contributed to a null surface by the matter crossing it. This, in turn, suggests looking for a corresponding heat density $\mathcal{H}_g$ contributed by gravity, such that extremizing the total heat density will lead to the relevant field equations. As we saw, it is indeed possible to construct such a thermodynamic variational principle \textit{not only} for general relativity, \textit{but also} for all \LL\ models. The construction is based on the tensor $P^{ab}_{cd}$, which determines the entropy of horizons in the appropriate theory. Thus, one has a completely self-consistent thermodynamic variational principle for a large class of gravitational theories.

This approach also suggests that the standard geometrical variables must have a thermodynamic interpretation, and we should be able to recast the field equations themselves into a thermodynamic language. We illustrated these features in Section~\ref{sec:geotherm}. One finds that the time evolution of the spacetime metric is driven by the difference ($N_{\rm sur} - N_{\rm bulk}$) between the appropriately-defined surface and bulk degrees of freedom. Static spacetimes obey holographic equipartition in which $N_{\rm sur} = N_{\rm bulk}$, thereby leading to the equality of the number of degrees of freedom in the surface and bulk. All of these ideas work both on a spacelike surface and on a null surface. In the case of the latter, the field equations can also be re-written as a Navier--Stokes equation, which is probably the most direct connection between gravity and fluid dynamics. Further, just as in the case of normal matter, the equipartition condition allows us to identify the number density of microscopic degrees of freedom. We found that there are $A/L_P^2$ degrees of freedom, which can be associated with an area $A$.

The second part of the review (Sections~\ref{sec:gravheatden} and \ref{sec:kineticsast}) takes this analysis one level deeper. The challenge is to obtain the expression for $\mathcal{H}_g$ from more fundamental considerations. We find that if we switch to the description of the differential geometry in terms of the geodesic interval $\sigma^2(x,x')$ rather than the metric, then the combination $R_{ab}n^an^b$
where $n_a=\nabla_a\sigma$ occurs rather ubiquitously in several geometrical expressions. The most primitive of these are the volume ($\sqrt{g}d^4x$) and area measures ($\sqrt{h}d^3x$) related to an equi-geodesic surface. In classical differential geometry, these measures $\sqrt{g}$ and $\sqrt{h}$ vanish when the equi-geodesic surface shrinks to a point. Therefore, even though the expressions for $\sqrt{g}$ and $\sqrt{h}$ contain the combination $R_{ab}n^an^b$, it does not contribute in the appropriate limit and prevents us from `associating' an area or volume with an event.

This is, of course, just an indication that the degrees of freedom of spacetime will arise only when we introduce a quantum of area $L_P^2$. There is a fair amount of evidence that suggests that one of the effects of quantum gravity is to introduce a zero-point length $L_0$ in the spacetime, by modifying $\sigma^2 \to \sigma^2 +L_0^2$. When this idea is developed further, in terms of a renormalized spacetime metric, which we called the q-metric, something remarkable happens. The volume measure corresponding to the renormalized metric still vanishes when the equi-geodesic surface collapses to a point; but the area measure remains finite and contains the correct expression for the heat density when we take $L_0^2 = (3/4\pi)L_P^2$. Thus, this approach allows us to count the number density of atoms of spacetime, and, by comparing the result with the macroscopic theory, determines the value of $L_0$. We also have a fundamental reason as to why the area measures are more relevant than the volume measures, a feature that has been repeatedly noticed in the thermodynamics of horizons.

The description at this layer is more speculative than in the previous part, but, of course, the rewards are also significantly higher. One can compare this layer of description with the kinetic theory of gases, which recognizes the existence of atoms, but yet, works at scales where a continuum description is~possible. 
The central quantity in such a description, in the case of a gas, will be the distribution function $f(x^i,p_j)$, which will give the number of atoms of gas at an event $x^i$ with momentum $p_j$. The corresponding quantity for the spacetime is a function $f(x^i,n_j)$
where $n_j = \nabla_j \sigma$ is the tangent vector to the null geodesic at the event $x^i$. Since several null geodesics can emanate from a given event, this is analogous to the distribution function for a gas, which describes several particles with different momenta coexisting at a single event. In neither case can the spacetime event be truly infinitesimal, and one assumes the existence of some intermediate scales, so that a sufficiently large number of atoms (of either gas or spacetime) can be collectively described by a distribution function. In the case of normal matter, one can think of $f(x^i,p_j)$ as counting the number of (i) atoms, or (ii) microscopic degrees of freedom, or (iii) microstates available to the discrete entities, since they all differ only by a numerical factor. In the case of spacetime, it seems appropriate to think of $f(x^i,n_j)$ as counting the number of microstates of geometry at $x^i$ with an internal degree of freedom described (at some suitable limit) by a null vector $n_j$. (The broad picture is somewhat reminiscent of Wheeler's spacetime foam idea \cite{jw}, but it is difficult to make a connection in general with only macroscopic inputs; the few computations based on spacetime foam (e.g.,~\cite{remo}) that exist are model dependent.)

There are several open questions that this description raises, and their investigation will prove to be fruitful in taking the ideas further. 
The most crucial question (which has not been tackled so far in the emergent gravity paradigm) is the role of normal matter, which has been introduced through a conserved energy momentum tensor $T^a_b$. While the macroscopic physics did provide an interpretation of $T^a_b\ell_a\ell^b$, which we used to develop the ideas further, this term lacks a microscopic description at present.In fact, it is rather ironic that, in our approach, we get the gravitational sector as a relic from quantum gravity, but have no quantum or semi-classical description of matter!\footnote{An analogy with a gaseous system is the following: think of a gas, confined to a box with a piston and described by a distribution function $f(\mathbf{x},\mathbf{p})$ giving the number density of atoms. Using $f(\mathbf{x},\mathbf{v})$, one could compute not only the pressure exerted on the piston, but also the fluctuations in the pressure which acknowledges  the existence of atoms in the gas. Even though both the piston and the gas are made of atoms and interact with each other, we are now taking into account the discrete constituents of the gas, but not of the piston. The situation in which we recognize the discrete nature of spacetime, but borrow $T_{ab}$ from the classical theory, is roughly analogous.}

This is one issue in which the thermodynamic variational principle lags behind the usual action principle; in the latter, one has a uniform description in terms of the sum of the actions, $A_{\rm grav} + A_{\rm matt}$, and the extremum principle for the action is sanctioned by the quantum theory. The thermodynamic variational principles for normal systems, for example, the one for entropy $S(q_A)$, however, do not come from any path integral \textit{amplitude}, but instead from the fact that the probability for a configuration is proportional to $\exp S(q_A)$. This would suggest that the gravitational sector of the variational principle should have a similar probabilistic interpretation.

If we interpret $f(x^i,n_j)$ as related to number of microscopic states available to quantum geometry, then in the suitable limit, one can introduce a probability $P(x^i,n_a)$ for $n_a$ at each event $x^i$ and the partition function:
\begin{equation}
 e^{S(x^i)}\propto \int\mathcal{D}n_i P(x^i,n_a)\exp[\mu L_P^4 T_{ab}n^an^b]
\label{geoz}
\end{equation} 
where $\mu$ is a numerical factor of order unity. If we take: 
\begin{equation}
 P(x^i,n_a)\propto\exp [\mu f(x^i,n_a)] \propto \exp\left( - \frac{\mu L_P^2}{8\pi} R_{ab} n^an^b\right)
 \label{eqnx}
\end{equation} 
then the steepest-descent evaluation of \eq{geoz} will pick out the geometry determined by Einstein's equation with an arbitrary cosmological constant. (Further, the choice $\mu=1/4$ will allow $P$ to be interpreted as the number of microstates.) More generally, one can think of $P(x^i,n_a)$ to be such that it gives the 
correlator: 
$
 \langle n^an^b\rangle \approx(4\pi/\mu L_P^2) R_{ab}^{-1}
$
which facilitates writing the field equations in the form:
\begin{equation}
 2\mu L_P^4\ \langle \bar T_{ab} n^an^b\rangle\approx 2\mu L_P^4\ \langle \bar T_{ab}\rangle \langle n^an^b\rangle =1
\label{mach}
\end{equation} 
where $\bar T^a_b\equiv T^a_b-(1/2)\delta^a_b T$ and $\langle \cdots \rangle$ now indicates both expectation values for the quantum operator $\bar T_{ab}$, as well as a probabilistic averaging of $n^an^b$. 

Equation~(\ref{mach}) has a clear Machian flavor. We cannot set $\langle T_{ab}\rangle =0$ everywhere and study the resulting spacetime, since it will lead to $0=1$! \textit{Matter and geometry must emerge and co-exist together, suggesting a new perspective on cosmology.} 
If \eq{geoz} could be obtained from a systematic approach, we will have a nice way of describing the effect of the source on the geometry.
This will also throw light on the avoidance of classical singularities in quantum gravity, which is definitely indicated in any spacetime with a zero-point length. 
In all such approaches, one would consider $f(x^i,n_a)$ as a fundamental (pre-geometric) object; from this point of view, it would be also interesting to study the evolution equation for $f(x^i,n_a)$ in terms of, say, $n^j\nabla_j f(x^i,n_a)$. 

The choice of $\sqrt{h}$ as a measure of the density of the atoms of spacetime seems reasonable, but one cannot ignore the fact that many other geometrical variables in the renormalized spacetime have \cite{D6} finite limits, containing the combination $R_{ab}n^an^b$, which is, in fact, rather ubiquitous. We have made the most basic choice, but it would be nice if one could explore other possibilities, as well. One possibility, for example, is the following: We know that in the local Rindler frame, $A_\perp/4$ is interpreted as entropy. We can compute the corrections to $A_\perp$ due to the curvature in the 
Euclidean sector, by computing the corresponding quantity over a small circle in the $T_E,X$ plane. (This is not quite an equi-geodesic surface, as we have defined it, but a cross-section of it on the $T_E,X$ plane; however, the idea still works.) Classically we find that the correction does have the factor $[1-(\sigma^2/6)(R_{ab}n^an^b)]$ where $n^a$ is now confined to the $T_E,X$ plane, which, of course, does not contribute in the $\sigma\to0$ limit. Working out the same with the q-metric (now with $g_{ab}$ corresponding to a Riemann normal coordinate system boosted to a local Rindler frame), we will get the correct result. Therefore, one can also interpret the entropy density $(R_{ab}n^an^b)$ as corrections to $A_\perp/4$ in flat spacetime. One can also do a similar exercise \cite{D6} with the integral of the extrinsic curvature
$K/8\pi$ over a stretched horizon in local Rindler frame, which we know gives its heat content $\kappa A_\perp/8\pi$ in flat spacetime; but in this case, one needs to make some \textit{ad hoc} choices for the numerical factors to get the result.
Such attempts, \textit{viz}. interpreting our extra terms as curvature corrections to the standard expressions for entropy (which works only after adding the zero-point length), are rather unsatisfactory as first-principle approaches.

There is another natural geometrical quantity that contains $(R_{ab}n^an^b)$. The expression for $f(x^i,n_a)$ comes from the term in square brackets in \eq{hlimit} which, in turn and rather surprisingly, arises from the ratio of Van Vleck determinants in \eq{defns}, which has the leading order behavior:
\begin{equation}
 \frac{\Delta}{\Delta _{S}}=f(x^i,n_a)
=1-\frac{1}{6} L_{0}^{2} R_{ab}n^an^b
\end{equation} 
so one could have used this as an alternative definition for $f(x^i,n_a)$. This might be better for the probabilistic interpretation of $f$ in \eq{eqnx}. 

Finally, it will be interesting to ask how these ideas generalize to \LL\ models (for some related ideas, see \cite{entent}). The renormalization of a \LL\ theory will, of course, lead to a different expression for $q_{ab}$, the corresponding $S(\sigma^2)$ and, consequently, a different expression for $\sqrt{h}$. However, for consistency, we know that the final $f(x^i,n_a)$ must be the same with $R_{ab}$ replaced by $\mathcal{R}_{ab}$. It will be interesting to explore whether these notions work out for \LL\ models, as well.

\section*{Acknowledgments}

I thank Sumanta Chakraborty, Sunu Engineer, Dawood Kothawala and Kinjalk Lochan for several discussions and comments on the manuscript. My work is partially supported by the JC Bose fellowship of the Department of Science and Technology (DST) of the Government of India.

\end{document}